\documentclass[twocolumn]{aastex7}
\usepackage{graphicx}	
\usepackage{amsmath}	
\usepackage{orcidlink}

\shorttitle{miscibility and the effect on the cores of sub-Neptune planets}

\usepackage{hyperref}

\newcommand{\revisedtext}[1]{\textcolor{black}{#1}}

\begin{document}

\title{Differentiation, the Exception Not the Rule - Evidence for Full Miscibility in Sub-Neptune Interiors }

\correspondingauthor{Edward D. Young}

\author[0000-0002-1299-0801]{Edward D. Young}
\affiliation{Department of Earth, Planetary, and Space Sciences, University of California, Los Angeles, CA 90095, USA}
\email[show]{eyoung@epss.ucla.edu}

\author[0009-0005-1133-7586]{Aaron Werlen}
\affiliation{Institute for Particle Physics and Astrophysics, ETH Z\"{u}rich, CH-8093 Z\"{u}rich, Switzerland}
\email{awerlen@ethz.ch}

\author[0009-0001-6858-9849]{Sarah P. Marcum}
\affiliation{Department of Earth, Planetary, and Space Sciences, University of California, Los Angeles, CA 90095, USA}
\email{smarcum13@g.ucla.edu}

\author[0000-0003-3778-2432]{Lars Stixrude}
\affiliation{Department of Earth, Planetary, and Space Sciences, University of California, Los Angeles, CA 90095, USA}
\email{lstixrude@epss.ucla.edu}

\author[0000-0002-7078-5910]{Cornelis P. Dullemond}
\affiliation{Institut f\"{u}r Theoretische Astrophysik, Zentrum f\"{u}r Astronomie, Universit\"{a}t Heidelberg, D-69120 Heidelberg, Germany}
\email{dullemond@uni-heidelberg.de}

\begin{abstract}
We investigate the consequences of non-ideal mixing between silicate, iron metal, and hydrogen for the structures of the cores of sub-Neptunes with implications for super-Earths and perhaps other planets formed by mixtures of silicate melt, iron metal, and significant mass fractions of hydrogen. A method of extrapolating what we know about the miscibility in the three bounding binary systems MgSiO$_3$-H$_2$, MgSiO$_3$-Fe, and Fe-H$_2$ to the ternary composition space is used to deduce the phase equilibria of this system at relevant temperature and pressure conditions.  We find that while separate silicate and metal phases can exist at shallow depths, the phases become entirely miscible deeper in the cores, thus altering the density structure of the cores. The assumption that the interiors of large rocky planets, either with extant magma oceans beneath H$_2$-rich envelopes, or evolved from such bodies, are composed of a differentiated metal core overlain by a silicate mantle is inconsistent with our understanding of the phase equilibria of these bodies.       
\end{abstract}


\section{Introduction}
Based on the Kepler survey, sub-Neptunes are the most abundant planet type in the Galaxy  \citep[e.g.,][]{fressin2013a, fulton2017a}.  These bodies have bulk densities, $\rho$, similar to about 1 to 3 g cm$^{-3}$ compared with Earth's bulk density of $5.5$ g cm$^{-3}$. Studies of the atmospheres of sub-Neptunes are multiplying rapidly in the age of the James Webb Space Telescope (JWST). However, our understanding of their cores, meaning everything beneath the outer envelope, is necessarily limited.  In some senses, we are in a period of discovery for sub-Neptunes that parallels our understanding of the core of the Sun in the 1920s \citep{Payne_1925}. We need judicious use of physical chemistry to infer what is beneath the outer veils of these copious planets.   

Some interpretations of their low densities relative to compressed rock and metal suggest that sub-Neptunes may be mainly a mixture of roughly equal mass fractions of rock and water \citep[e.g.,][]{Fortney2007, zeng2019a}. To see this, consider that if a sub-Neptune with a density of 2 g cm$^{-3}$ were to be composed of water, with $\rho = 1$ g cm$^{-3}$, and a core with a bulk compressed core density of $5.5$ g cm$^{-3}$, similar to that of Earth, it would have a mass fraction of water of about 40\%.  The sources of this putative water are debated  \citep{Bitsch_2019}. Objects in the Kuiper Belt provide one model. Assuming Pluto is composed of a mix of water ice and chondrite-like rock, it has a mass fraction of water of about 20\%, suggesting that Kuiper belt objects are not sufficiently water-rich to provide the feedstock for the proposed water worlds. 

Alternatively, sub-Neptunes may have rock-like, molten or partially molten interiors beneath hydrogen-rich primary atmospheres \citep{Chachan2018}. The H$_2$-rich envelopes in this interpretation must comprise percent levels of the planet masses in order to match their bulk densities \citep[e.g.][]{Bean2021}. For example, assuming a density for H$_2$ of 0.1 g cm$^{-3}$, appropriate for the high pressures and temperatures at the base of the envelopes \citep[e.g.,][]{Chabrier2019}, a compressed core density similar to that for the whole Earth, and a median sub-Neptune density of 2 g cm$^{-3}$, the planet would have an H$_2$ envelope comprising $2.7$ \% of its  mass.  

The dense envelopes in this case are sufficient to delay cooling of the originally hot planets such that the magma oceans beneath may persist for Gyr timescales, depending on the processes of atmospheric escape attending the evolution of each planet \citep{Vazan2018, ginzburg2016a, Rogers_2024}. This implies that extant magma oceans are present among many, and perhaps even most, sub-Neptunes observed today. 

There is some observational support for the magma ocean-hydrogen-rich envelope structure for sub-Neptunes. The JWST spectrum for the atmosphere of sub-Neptune K2-18 b \citep{Benneke_2019}, touted as a water-rich ``Hycean" planet \citep{Madhusudhan_2020}, can be explained by equilibration between a hydrogen-rich envelope and an underlying magma ocean \citep{Shorttle_2024}. The metal-rich, relatively low C/O atmosphere of sub-Neptune GJ 3090 b could be attributed to interaction with an underlying magma ocean, although a plausible alternative is that the envelope chemistry has been modified by atmospheric escape \citep{Ahrer_2025}. The outer envelope of sub-Neptune TOI-270 d also exhibits chracteristics that can be explained by interaction with an underlying magma ocean, including a mean molecular weight of $5 \pm 1$ amu \citep{Benneke2024,Felix2025}. 

The bulk densities of super-Earths indicate that they are rocky planets with broadly Earth-like bulk densities and negligible atmospheres in terms of mass. A strong case can be made that super-Earths form when some sub-Neptunes lose their dense hydrogen-rich envelopes \citep{Rogers_2024}, leaving behind rocky cores. This leads to the well-known radius gap that separates these two classes of planets \citep{fulton2017a, owen2013a, gupta2019a}: super-Earths have radii between $>1$ and $1.7 R_\oplus$, while sub-Neptunes range from about $1.9$ to $4.0R_\oplus$. In this view, super-Earth interiors may be similar to those of sub-Neptunes \citep{Bean2021}. An important caveat is the prospect for degassing of H$_2$ from the molten interior of a sub-Neptune as it loses its envelope and cools inexorably below some threshold final equilibration temperature \citep[e.g.,][]{Rogers_2025}.  While super-Earths appear to evolve from sub-Neptunes, it is not yet clear that they retain all of the properties of sub-Neptune cores.  

It has been suggested recently that the interface between sub-Neptune magma ocean cores and H$_2$-rich envelopes should correspond to a phase change from a supercritical magma ocean composed of silicate and hydrogen, to molten silicate and a separate hydrogen-rich phase \citep{Markham2022,Young_2024}.  This suggestion redefines the meaning of the magma ocean - envelope interface.  More specifically, the binodal (sometimes referred to as a coexistence curve, or solvus) in the MgSiO$_3$-H$_2$ system in this context corresponds to the temperature at a given pressure where a supercritical mixture of silicate and hydrogen exsolves to form two discrete phases \citep{Stixrude2025}.  Identifying the surface of magma oceans with a first-order phase transition at a binodal underscores the fact that phase equilibria at relevant pressure and temperature conditions can lead to quite different pictures of these planets relative to our familiar reference points in the Solar System; sub-Neptunes are not simply scaled-up Earths with primary hydrogen atmospheres \citep{Young_2024}.     

Here we investigate the nature of the molten cores of sub-Neptunes, and by inference the cores of their super-Earth descendants and perhaps even the deep interiors of their warm Neptune and ice giant cousins.  In particular, we investigate whether sub-Neptunes likely have iron-rich cores like the terrestrial planets in our Solar System.  This question is important because it is common to model the cores of these planets as if they differentiated into Fe-rich metal cores (senso stricto) and silicate mantles \citep[e.g.,][]{Davenport_2025}, based on analogy with the rocky bodies in the Solar System.  The nature of the cores (senso lato, everything other than the envelope) is relevant to estimate the masses of the hydrogen-rich envelopes required to explain their bulk densities. The more hydrogen the cores retain, the less the density deficits relative to pure rock can be attributed to the outer envelope \citep[e.g.,][]{Kite2019, Schlichting_Young_2022}. 

Our paper is structured as follows. In section \ref{Prospects} we discuss bulk chemical constraints on iron-rich cores for exoplanets in general.  In section \ref{Mixing_models} we describe the binary mixing relationships that are the basis for our analysis. In this section we present existing and newly-derived binary Gibbs free energy surfaces, and we present three new {\it ab initio} - molecular dynamics simulations in support of these free energy surfaces. In section \ref{Ternary_mixing}  we present our new  ternary Gibbs free energy surface based on the binary systems, and a convex hull approach for calculating the stable phase equilibria in the ternary system.  The resulting ternary phase diagram is presented in section \ref{Results} and implications for the interior structures of sub-Neptunes are discussed in section \ref{Discussion}.  Conclusions are presented in the final section.

\section{Prospects for extrasolar Fe-rich metal cores}
\label{Prospects}
\subsection{Bulk chemistry}
A first-order constraint on whether or not sub-Neptunes, as well as other extrasolar rock-rich planets, can harbor metal-rich cores comes from their bulk chemical compositions as inferred by stellar elemental abundance ratios. Correlations between rocky planet bulk densities and the relative abundances of Fe, Mg, and Si in their host stars yield inferred mass fractions of metal cores of rocky exoplanets in the range of 25 to 35\% \citep{Adibekyan_2021}. The assumption that most iron resides in metal is validated by observations of polluted white dwarfs (WD). The iron concentrations in silicate-dominated materials accreted by WDs indicate that they have intrinsic oxygen fugacities similar to planetary building blocks in the Solar System \citep{doyle2019a}. This in turn suggests that the planets built from these materials could have Earth-like, iron-rich metal fractions \citep{Trierweiler_2023}. There is also some more direct observational evidence for rock-metal differentiation for a body orbiting around a WD. \cite{Manser_2019} reported periodic shifts in emission spectral lines from an accretion disk around a white dwarf that are explained by a remnant metal core of a differentiated planetesimal orbiting within the disk around the WD. The metal-like density of the spherical body has apparently permitted the body to avoid tidal disruption as it orbits so close to the WD. The bodies that pollute WDs are generally agreed to be similar to the largest asteroids in the Solar System asteroid belt \citep{Jura_2014, Trierweiler_2022}. Evidence for iron metal polluting a WD is thus evidence for silicate-metal differentiation at relatively low pressures and temperatures but not under conditions relevant for sub-Neptunes. 

\subsection{Physical differentiation}
The bulk chemical compositions of extrasolar rocky bodies or their source materials appear to be consistent with the formation of iron-rich metal cores. The issue then becomes whether the physics associated with differentiation, and the phase equilibria in the silicate-iron-hydrogen system, are such that silicate-metal differentiation occurs in the depths of sub-Neptunes.  

There is reason to believe that metal cores are not the norm for sub-Neptunes, and possibly not for their super-Earth descendents. Magma oceans beneath dense hydrogen-rich envelopes can persist for  $10^9$ yrs under favorable circumstances \citep{ginzburg2016a, Rogers_2024}. Thermodynamic simulations of the global equilibration between Fe-rich metal, silicate (represented by MgSiO$_3$), and H$_2$-rich envelopes show that metal should have significant fractions of H, O, and Si \citep{Schlichting_Young_2022}. As a result, metal droplets with sizes controlled by their integrity against shear stresses can become neutrally buoyant, forestalling formation of a metal core \citep{Young_2024}. Similarly, \cite{Lichtenberg2021b} concluded that turbulent entrainment of metal droplets is also likely simply because of the great depth of the sub-Neptune or super-Earth magma oceans. While it is important to recognize that complete physical separation between metal and silicate can be questioned in the case of sub-Neptunes, this is not the focus of this paper. 

\subsection{Miscibility}
The focus of this paper is on the phase equilibria of the MgSiO$_3$-Fe-H$_2$ system and implications for the phases present in the deep interiors of sub-Neptunes. There is evidence from {\it ab initio} molecular dynamics simulations, as well as experiments,  that deep in the interiors of sub-Neptune magma oceans, discrete silicate and iron-rich metal phases do not exist.  These constraints include the existence of complete miscibility between molten MgO and Fe at temperatures above 7000 K and pressures of 60 GPa \citep{Wahl_2015, Insixiengmay_2025} and extensive solubility along this binary join at lower temperatures and similar pressures \citep{Badro_2018}, complete miscibility between liquid MgSiO$_3$ and H$_2$ above about 4000 K at, for example, 4 GPa \citep{Stixrude2005}, and a pressure-dependent enhancement in the solubility of H in Fe melt relative to silicate melt \citep{Tagawa_2021, Yuan_2020}.  All of these studies point to the possibility that silicate and iron metal are miscible. We test this proposition here.

\section{Binary mixing models}
\label{Mixing_models}
\subsection{General approach}
In a closed system in which the composition of the system, specified by a set of mole fractions $x_i$ for components $i$, temperature $T$, and pressure $P$, are constant, phase diagrams can be constructed by minimizing the molar Gibbs free energy, $\hat{G}$. Gibbs free energy is the relevant Legendre transform of internal energy where the variables of interest are composition, $T$ and $P$.  We focus on the Gibbs free energy of mixing, $\hat{G}_{\rm mix}$, to determine the number and compositions of phases at specified conditions. Toward this end, we make use of two curves in parameter space that are derived from the free energy of mixing. In a binary composition space, the binodal is the loci of compositions for two coexisting stable phases. The spinodal lies interior to the binodal and is the region in composition-$T$-$P$ space where spontaneous transformation from one phase to two phases is predicted. Between these curves is a region of metastability where two phases may have metastable compositions. One can think of the spinodal as defining the region in parameter space where two phases are assured, and the binodal as separating the regions of one-phase and two-phase stability at thermodynamic equilibrium.

\subsection{\texorpdfstring{MgSiO\textsubscript{3}-H\textsubscript{2} system}{MgSiO3-H2 system}}

\cite{Stixrude2025} explored the non-ideal mixing between MgSiO$_3$ and H$_2$ using density functional theory molecular dynamics (DFT-MD) simulations, including relevant thermodynamic parameters. They find that the system becomes entirely miscible at temperatures greater than approximately 4000 K at relevant pressures. 
There is a negative dependence of the peak temperature for the crest of the binodal on pressure. The crest of the binodal (solvus) is near 3000 K at 10 GPa, and 4000 K at 2 GPa. We adopt their subregular mixing model without modification. Their fit to their DFT-MD simulations yields the free energy of mixing expression  

 \begin{equation}
 \begin{aligned}
\Delta \hat{G}_{\text{mix}} =& \left(L_{\mathrm{CB}} x + L_{\mathrm{BC}} (1-x)\right) x (1-x) \left( 1 - \frac{T}{\tau} + \frac{P}{\pi} \right) + \\
&R T \left( x \ln x + (1-x) \ln (1-x) \right).
\label{eqn:MgSiO3_H2}
 \end{aligned}
 \end{equation}
Here, $L_{\mathrm{CB}}$ and $L_{\mathrm{BC}}$ are the binary interaction parameters where the logic for the labels CB and BC will be apparent when we apply these to our ternary model, and $x$ is the mole fraction of H$_2$ along the binary join. In this formulation, the excess, or non-ideal, molar enthalpy of mixing is $\Delta\hat{H}_\mathrm{excess} = \left(L_{\mathrm{CB}} x + L_{\mathrm{BC}} (1-x)\right) x (1-x)$.  The temperature and pressure dependence for the non-ideal mixing is embodied by excess entropy and volume parameters, represented by $\tau$ and $\pi$ that are $\Delta\hat{H}_\mathrm{excess}/\hat{S}_\mathrm{excess}$ and $\Delta\hat{H}_\mathrm{excess}/\Delta\hat{V}_\mathrm{excess}$, respectively \citep{Stixrude2025}. The values for the subregular solution fit to the DFT simulations are:  $L_{\mathrm{CB}} = 622000$ J/mol,  $L_{\mathrm{BC}} = -4950$ J/mol, $\tau = 4800$ K,  and $\pi = -35$ GPa. 

\begin{figure}
\centering
   \includegraphics[width=0.45\textwidth]{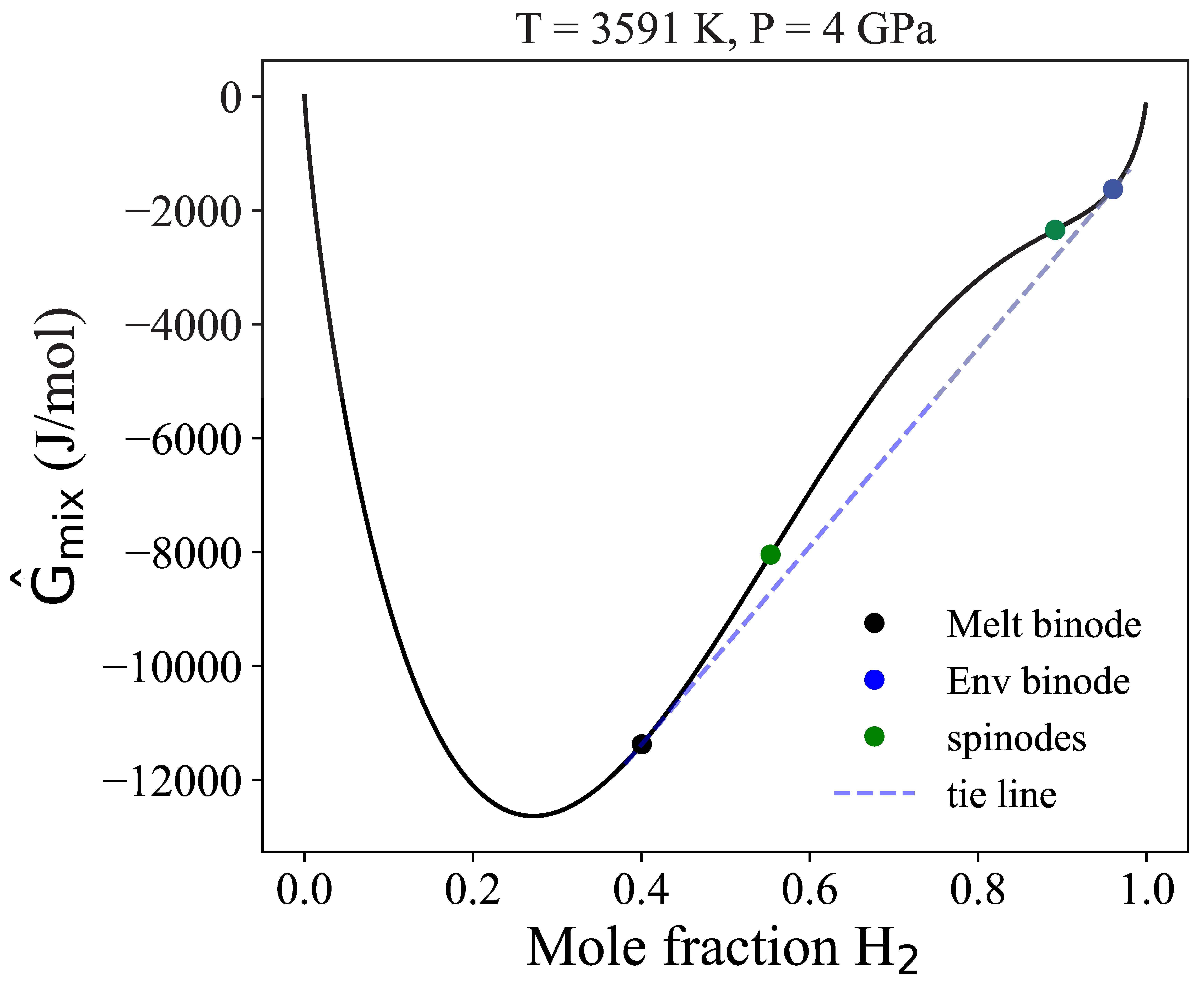}
    \caption{The Gibbs free energy of mixing surface for the MgSiO$_3$-H$_2$ system at 3591 K and 4 GPa as formulated by \cite{Stixrude2025}.  The binodes for the melt and envelope phases at this $T$ and $P$, as well as the tie line between them, are shown for clarity.  Also shown are the positions of the spinodes where the curvature is zero.}
\label{fig:Gmix_2D}
\end{figure}

\begin{figure}
\centering
   \includegraphics[width=0.45\textwidth]{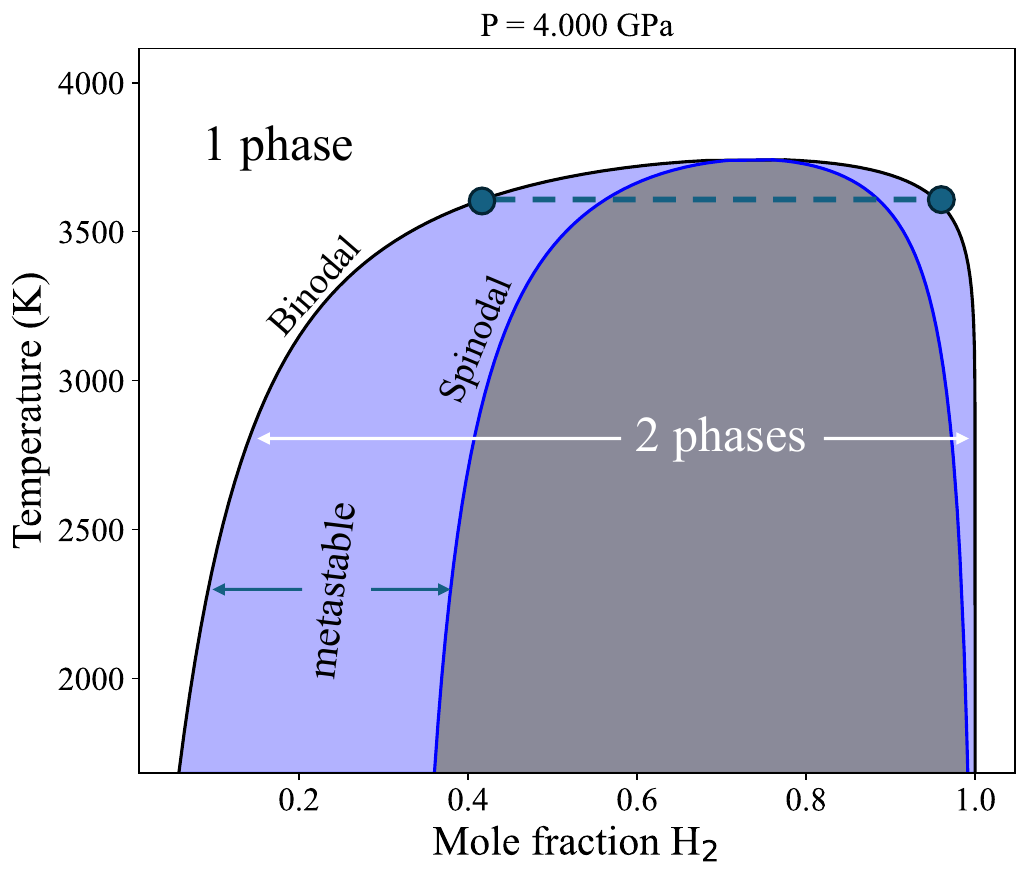}
    \caption{Isobaric phase diagram for the MgSiO$_3$-H$_2$ binary system as determined by \cite{Stixrude2025}. The 1 and 2-phase regions outside and inside of the binodal are indicated.  A region of metastable compositions for two coexisting phases exists between the binodal and spinodal curves. The compositions of two stable phases at approximately 3600 K are shown by way of example with a dashed tie line connecting them.}
\label{fig:isobaric_phase_diagram}
\end{figure}

The MgSiO$_3$-H$_2$ system can be used to illustrate the various aspects of the free energy of mixing that we make use of here. The Gibbs free energy of mixing surface for this binary system at 3591 K and 4 GPa is shown in Figure \ref{fig:Gmix_2D}. The binodes and spinodes are indicated. In Figure \ref{fig:isobaric_phase_diagram} we show the isobaric phase diagram derived from this free energy surface at different temperatures, where the complete binodal and spinodal curves are derived from points at a range of temperatures like those shown in Figure \ref{fig:Gmix_2D} for one temperature.

\begin{figure}
\centering
   \includegraphics[width=0.44\textwidth]{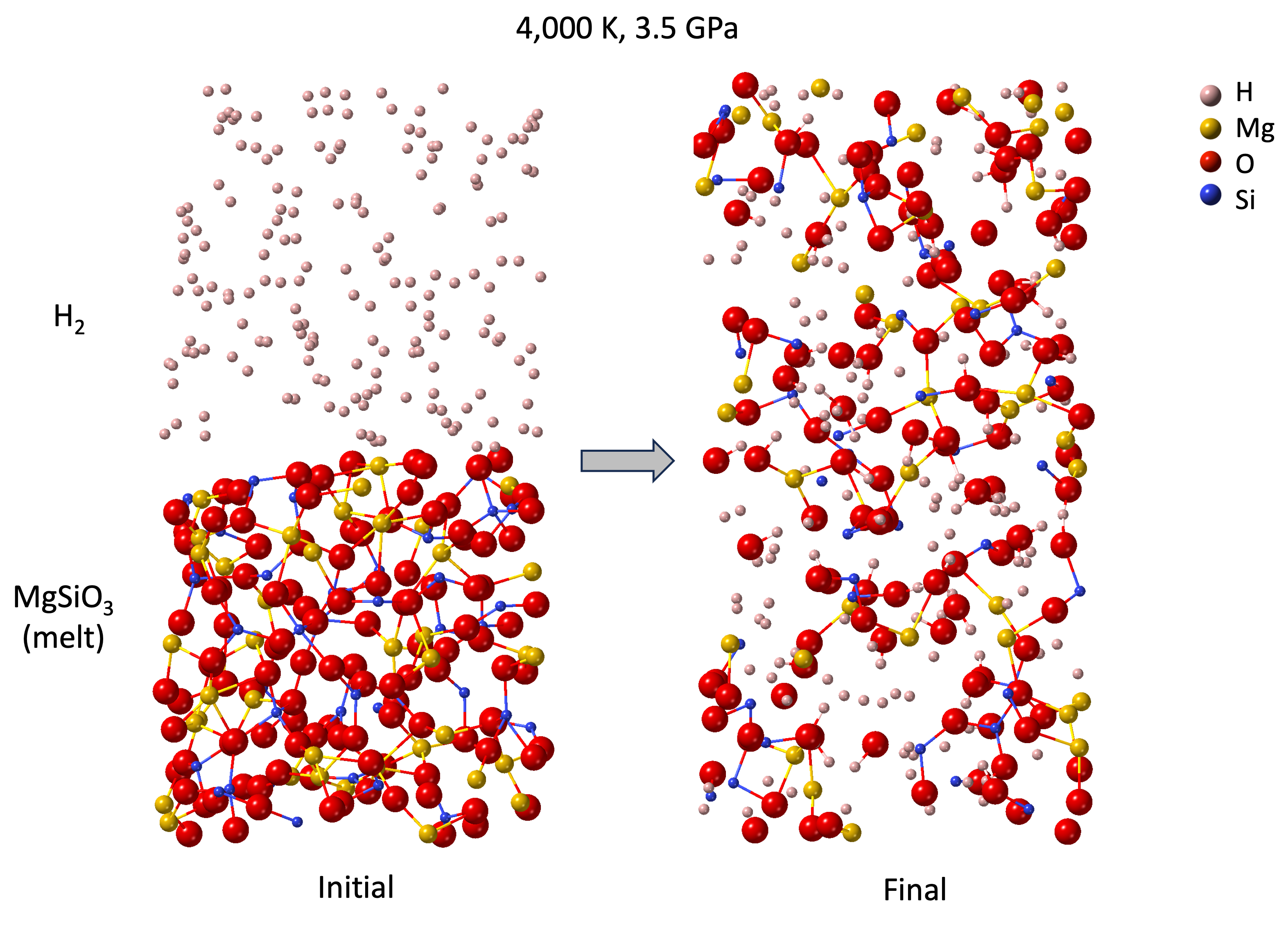}
    \caption{Images of initial conditions and final conditions for the H$_2$ - MgSiO$_3$ system illustrating complete miscibility at 4,000 K and $3.5$ GPa. The initial condition is composed of MgSiO$_3$ melt overlain by H$_2$ gas.  The final state is after $13.5$ ps of model time.  See Appendix B for details.  }
\label{fig:4000K_miscibility}
\end{figure}

\begin{figure}
\centering
   \includegraphics[width=0.44\textwidth]{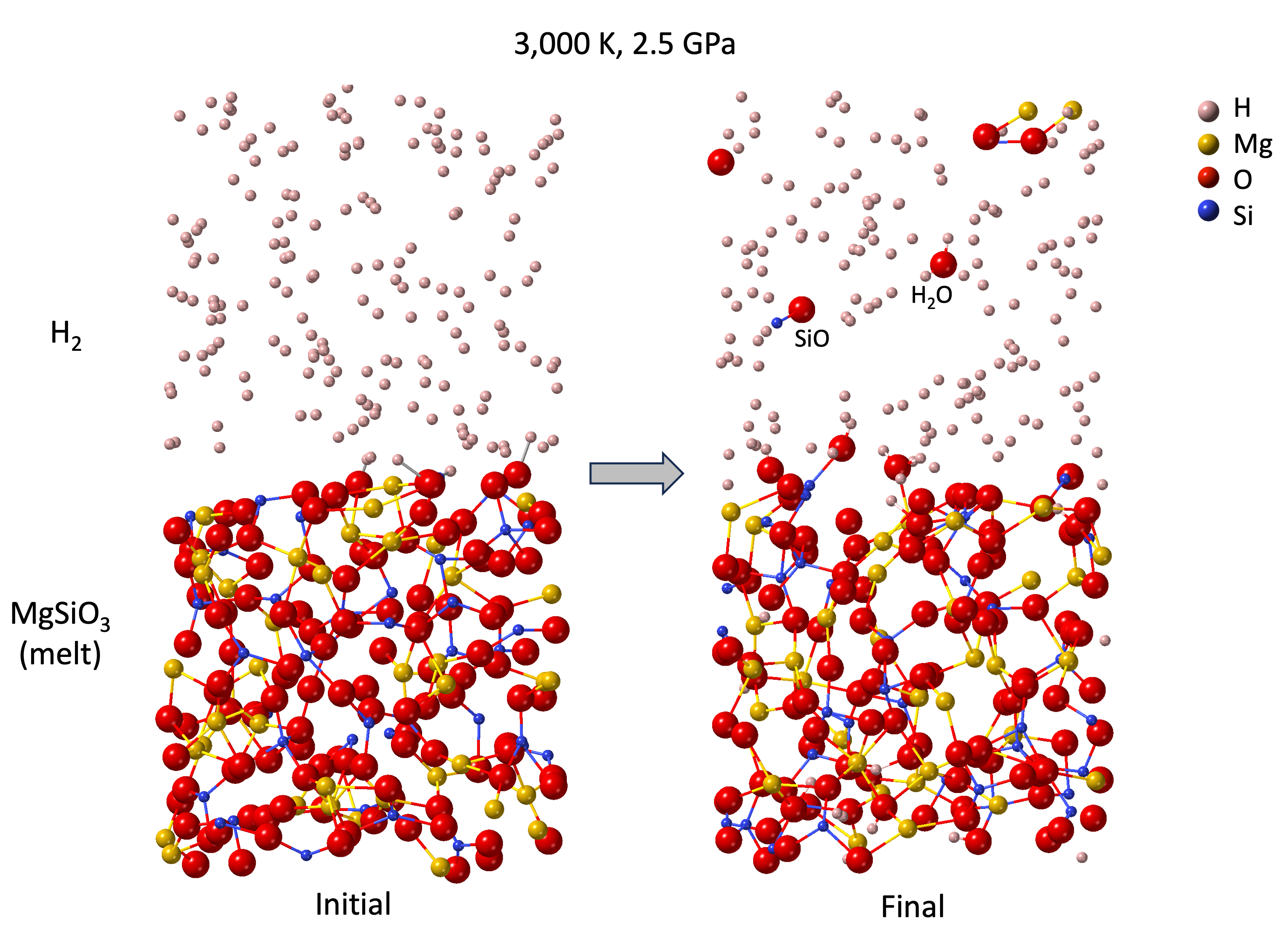}
    \caption{ Images of initial conditions and final conditions for the MgSiO$_3$-H$_2$ system illustrating immiscibility at 3,000 K and $2.5$ GPa. The initial condition is composed of MgSiO$_3$ melt overlain by H$_2$ gas. The final state is after $16.5$ ps  of model time.  Example SiO and H$_2$O molecules in the gas phase that persist for significant intervals of model time are labelled.  }
\label{fig:3000K_imiscibility}
\end{figure}

Here we present two new DFT-MD simulations that bracket the position of the binodal (Figures \ref{fig:4000K_miscibility} and \ref{fig:3000K_imiscibility}) as a check on the veracity of the model and an illustration of the two-phase mixing calculations.  Methods for the DFT-MD calculations are described in Appendix A. The initial condition for these simulations is composed of two distinct, adjacent phases, MgSiO$_3$ melt and H$_2$ gas, at the same temperature and about the same pressure (within  about 5\%). The hydrogen in each case comprises 4\% by weight of the system, corresponding to a composition near the peak of the binodal. After suitably long simulation times of $> 13$ ps, results show that the silicate-H$_2$ system is completely miscible at 4,000 K and $3.5$ GPa (Figure \ref{fig:4000K_miscibility}), but not at 3,000 K and $2.5$ GPa (Figure \ref{fig:3000K_imiscibility} ), where the different pressures were used to preserve number of atoms and volumes for the two different temperatures.  This shows that the implied $T$ and $P$ for the binodal for this bulk composition are indeed near those expected for  sub-Neptunes with of order a few weight per cent H$_2$. 

\subsection{MgSiO\texorpdfstring{$_3$}{3}-Fe system}
There are two published studies showing that molten MgO and Fe become completely miscibile at temperatures and pressures relevant for the interiors of sub-Neptunes.  \cite{Wahl_2015} used DFT molecular dynamics calculations and showed that at low pressures, the closure of the binodal (solvus) can occur at temperatures as low as 4000 K with a positive pressure dependence. The non-ideal mixing is slightly asymmetrical with the crest of the binodal being about 6000 K at 50 GPa, 7000 K at 100 GPa, and 9000 K at 400 GPa.  

More recently, \cite{Insixiengmay_2025} used DFT molecular dynamics to show that liquid MgO and Fe become completely miscibile above a temperature of 7000 K at a pressure of 60 GPa and above 9000 K at 145 GPa. These authors fit their simulations to a regular (symmetric) solution model. Their model can be represented as  

\begin{equation}
\begin{aligned}
\Delta \hat{G}_{\text{mix}} &= L_{\mathrm{AC}}(T,P) x (1-x)  +\\
&R T \left( x \ln x + (1-x) \ln (1-x) \right),
\end{aligned}
\label{eqn:MgSiO3-Fe}
\end{equation}
where $L_\mathrm{AC} = L_{\mathrm{CA}} = 240000 -28T\mathrm{(K)} + 1116 P\mathrm{(GPa)}$ (J/mol).

To our knowledge there are no published analogous mixing calculations for MgSiO$_3$-Fe. Nonetheless, 
it is reasonable to infer that there is non-ideal mixing along this join, as in the case of MgO-Fe, and that there is a well-defined binodal that closes at sufficiently high temperatures and pressures.  

\begin{figure}
\centering
   \includegraphics[width=0.48\textwidth]{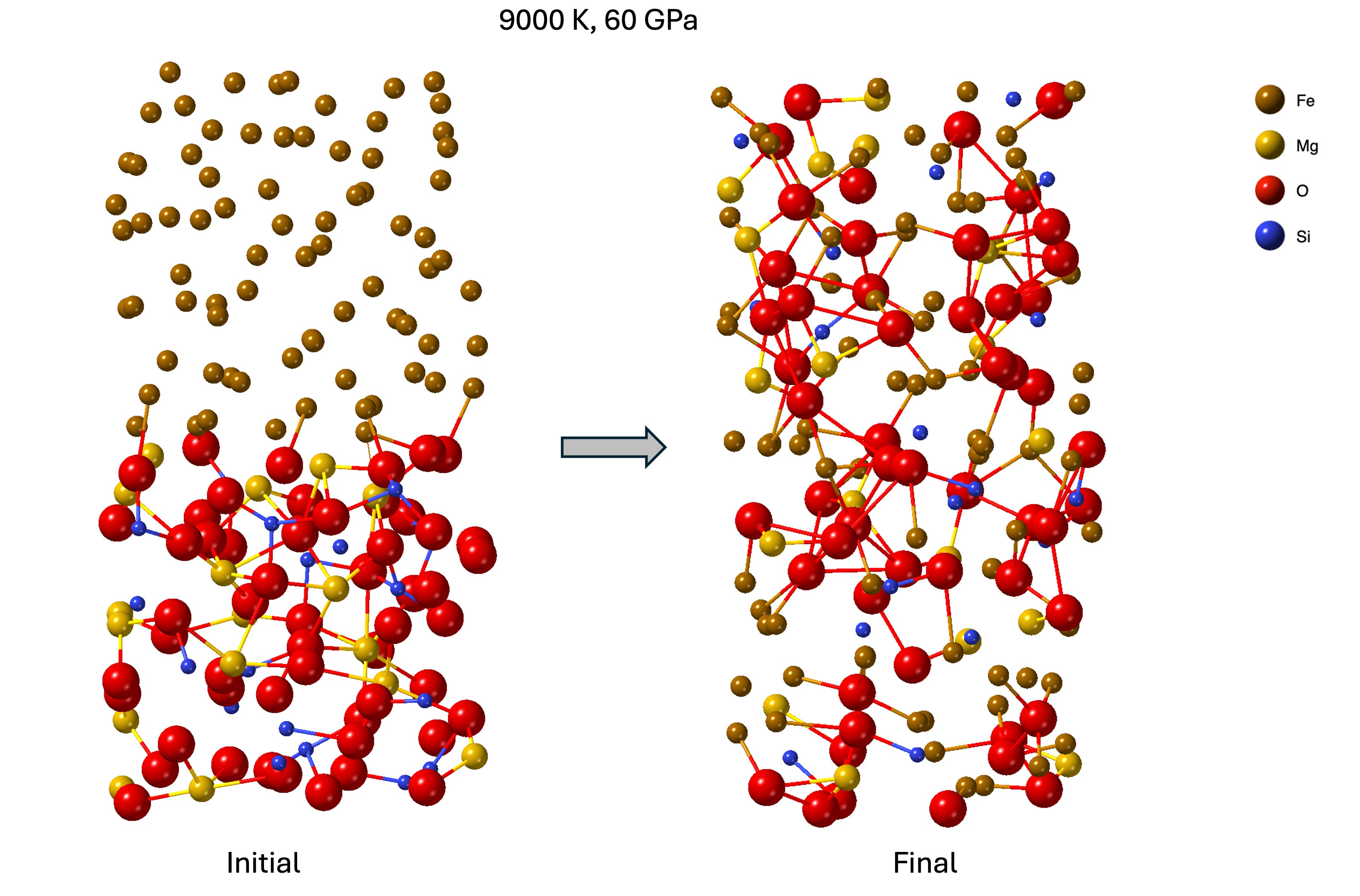}
    \caption{Images of initial conditions and final conditions for the  MgSiO$_3$-Fe system illustrating complete miscibility at 9,000 K and $60$ GPa. The final state is after 7 picoseconds of model time. Time steps were 0.5 fs.  Model consists of 76 Fe atoms, 18 Mg atoms, 18 Si atoms, and 54 O atoms.}
\label{fig:9000K_miscibility}
\end{figure}

We performed DFT-MD simulations of the MgSiO$_3$-Fe system in order to test whether the regular solution model for MgO-Fe mixing can be used as a surrogate for the MgSiO$_3$-Fe system at high $T$ and $P$. Details of the calculations are presented in Appendix A. Results show that at conditions above the MgO-Fe binodal, MgSiO$_3$ and Fe are indeed completely miscibile (Figure \ref{fig:9000K_miscibility}).  

There is experimental support for miscibility of MgSiO$_3$ and Fe at high $T$ and $P$.  \cite{Badro_2018} observed that the mole fraction of Fe in metal equilibrating with silicate at 4000 K and 54 GPa is only $0.65$, with the remainder of the metal phase composed of Mg, Si and O in roughly 1:1:3 proportions.  Similar results were obtained for Fe, Si, Mg, and O by \cite{Shakya_2024} using machine learning methods for molecular dynamics simulations at 3000 to 4000 K and 30 to 35 GPa. This level of mutual solubility at these temperatures and pressures is in reasonable agreement with the DFT calculations. 

\subsection{Fe-H\texorpdfstring{\textsubscript{2}}{2} system}
Constraints on the thermodynamics of mixing between Fe and H$_2$ at relevant $T$ and $P$ come from experiments and {\it ab initio} modeling.  At relatively low temperatures and pressures (e.g., 2000 K, 0.1 GPa), iron-rich melt coexists with H$_2$ gas \citep{SanMartin_1990}.  There is, however, both experimental and modeling evidence that the solubility of H$_2$ in liquid iron rises sharply at high $T$ and $P$. \cite{Tagawa_2021} performed experiments measuring the distribution coefficient for H in molten Fe metal relative to H in molten silicate, $D_\mathrm{H} = [H]_\mathrm{metal}/[H]_\mathrm{silicate}$, from 3100 to 4600 K and 30 to 60 GPa. They found that while $D_\mathrm{H}$ is near unity at $T < 3000$ K, it increases to 10 at 3000 K and 30 GPa and to 30 at 4000 K and 60 GPa. Similarly, \cite{Yuan_2020} performed {\it ab initio} modeling showing that $D_\mathrm{H}$ increases from near unity at 0 GPa to about 100 at 40 GPa. These observations all point to an enhancement in the solubility of H$_2$ in molten Fe metal with pressures greater than about 10 GPa. The requirement for progressive closure of the miscibility gap between H$_2$ and liquid iron with pressure \cite{Fukai_1992}, and the relative distribution of H between molten silicate and iron (increasing $D_\mathrm{H}$) permit a rough calibration of the Fe-H$_2$ mixing relations. 

An additional constraint comes from the parameterization of the solubility of H in molten Fe at lower temperatures and pressures.  The compilations by \cite{ZHANG_2007_FeH2} and \cite{Jiang_2011} yield the expression 

 \begin{equation}
 \begin{aligned}
\ln(x_{\rm H}) = -4.423 - \frac{4009 {\rm K}}{T} + \frac{1}{2} \ln\left(\frac{P_{\rm H_2}}{\rm 1 bar}\right).
\label{eqn:Fe_H2_low_PT}
 \end{aligned}
 \end{equation}
We make use of this Equation as a constraint on the Gibbs free energy surface for mixing of molten Fe and H$_2$ by matching the melt limb of the binodal to the solubility expressed by Equation \ref{eqn:Fe_H2_low_PT} at conditions appropriate for our fiducial sub-Neptune magma ocean-H$_2$-rich envelop interface (the minimum required extrapolation of $T$ and $P$).

\begin{figure}
\centering
\includegraphics[width=0.45\textwidth]{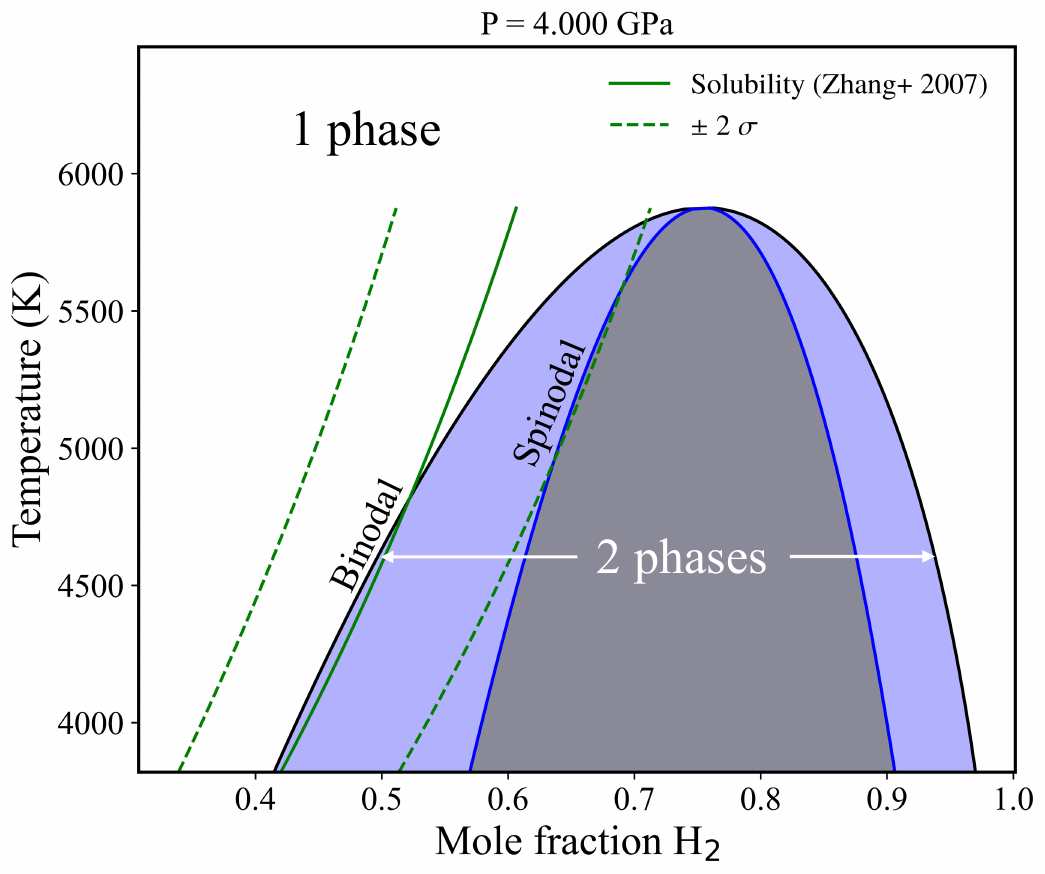}
\caption{Isobaric phase diagram for the Fe–H$_2$ binary system based on Equation \ref{eqn:Fe_H2}. The extrapolation of the low-$P$ and low-$T$ hydrogen solubility in liquid Fe metal is shown for comparison \citep{ZHANG_2007_FeH2,Jiang_2011}. Errors for the latter are derived by standard propogation of uncertainties.}
\label{fig:isobaric_phase_diagram_Fe_H2}
\end{figure}

We find that all of these relationships are reproduced using a subregular solution model where the interaction parameters are pressure dependent. The model is

 \begin{equation}
 \begin{aligned}
\Delta \hat{G}_{\text{mix}} =& \left(L_{\mathrm{AB}} x + L_{\mathrm{BA}} (1-x)\right) x(1-x)+ \\
&R T \left( x \ln x + (1-x) \ln (1-x) \right),
\label{eqn:Fe_H2}
 \end{aligned}
 \end{equation}

\noindent where  $L_{\mathrm{AB}} = 138000 -9500 P(\mathrm{GPa})$ J/mol, $L_{\mathrm{BA}} = 17000 -9500 P(\mathrm{GPa})$ J/mol, and $x$ is the mole fraction of H$_2$ on the binary join. The resulting isobaric phase diagram at 4 GPa is shown in Figure \ref{fig:isobaric_phase_diagram_Fe_H2}.  While the precise values used here are certainly not unique, the general shape of the free energy of mixing surface is required by the observations described above.  

\section{Ternary mixing model}
\label{Ternary_mixing}
We calculate the topology of the Gibbs free energy of mixing in the ternary system MgSiO$_3$-Fe-H$_2$ as the basis for our analysis.  
We make use of the mixing models for the three binary systems MgSiO$_3$-H$_2$, Fe-H$_2$, and MgSiO$_3$-Fe and extrapolate these bounding binary mixing models into the ternary system. There is a long history of performing such extrapolations with success with percent levels of accuracy \citep{Wohl_1946, Wohl_1953, Muggianu_1975, Jacob_1977, Lee_Kim_2001}. We adopt the Muggianu-Jacob method of projecting along shortest distances from each of the binaries to any given point in the ternary system \citep{Muggianu_1975, Jacob_1977}.

Labeling the three mole fractions comprising our mixture of Fe, H$_2$, and MgSiO$_3$ as $x_{\rm A}$, $x_{\rm B}$, and $x_{\rm C}$, respectively,  where $x_\mathrm{A}+x_\mathrm{B}+x_\mathrm{C} =1$, the ideal free-energy of mixing is

\begin{equation}
\begin{aligned}
\Delta\hat{G}_{\text{mix,ideal}} &= \\
& R T \left( x_{\mathrm{A}} \ln x_{\mathrm{A}} + x_{\mathrm{B}} \ln x_{\mathrm{B}} + x_{\mathrm{C}} \ln x_{\mathrm{C}} \right).
\end{aligned}
\label{eqn:ideal_mixing}
\end{equation}
Equation \ref{eqn:ideal_mixing} accounts for the ideal entropy of mixing.  
The excess, or non-ideal, free energy of mixing based on the subregular solution formulation is \citep{Ganguly_2001}:

\begin{equation}
\begin{aligned}
&\Delta \hat{G}_{\text{excess}} = \\
& x_{\mathrm{A}} x_{\mathrm{B}} \left( L_{\mathrm{AB}}  \frac{1}{2} (1 + x_{\mathrm{B}} - x_{\mathrm{A}}) + L_{\mathrm{BA}}  \frac{1}{2} (1 + x_{\mathrm{A}} - x_{\mathrm{B}}) \right) \\
& + x_{\mathrm{B}} x_{\mathrm{C}} \left( L_{\mathrm{BC}}  \frac{1}{2} (1 + x_{\mathrm{C}} - x_{\mathrm{B}}) + L_{\mathrm{CB}}  \frac{1}{2} (1 + x_{\mathrm{B}} - x_{\mathrm{C}}) \right) \\
& (1 - T/\tau + P/\pi) \\
& + x_{\mathrm{C}} x_{\mathrm{A}} \left( L_{\mathrm{CA}}  \frac{1}{2} (1 + x_{\mathrm{A}} - x_{\mathrm{C}}) + L_{\mathrm{AC}}  \frac{1}{2} (1 + x_{\mathrm{C}} - x_{\mathrm{A}}) \right) \\
& + x_{\mathrm{A}} x_{\mathrm{B}} x_{\mathrm{C}} L_{\mathrm{ABC}},
\end{aligned}
\label{eqn:Gexcess}
\end{equation}
where $L_{ij}$ are the interaction parameters for species $i$ and $j$, and $\tau$ and $\pi$ parameterize the temperature and pressure dependencies of the non-ideal mixing parameters for the MgSiO$_3$-H$_2$ subregular solution model \citep{Stixrude2025}. Expressions of the form $\frac{1}{2} (1 + x_i - x_j)$ are the normal projections of the mole fractions of the independent component $i$ onto the binary $i-j$ join. 

We set $L_{\mathrm{ABC}} = 0$ in Equation \ref{eqn:Gexcess} because it is unknown.  Indeed, deriving the ternary phase diagram in the absence of explicit values for ternary interaction parameters is the goal of using the Muggianu-Jacob projection method.  

\subsection{Spinodal}
The spinodal curve defines the region where spontaneous decomposition into two phases is required by the negative curvature of the free energy of mixing surface.  It  is therefore defined by the condition that the determinant of the Hessian matrix of $\hat{G}_{\text{mix}}$ with respect to composition vanishes, i.e.,

\begin{equation}
\det \left( 
\begin{bmatrix}
\frac{\partial^2 \hat{G}_{\mathrm{mix}}}{\partial x_{\mathrm{A}}^2} & \frac{\partial^2 \hat{G}_{\mathrm{mix}}}{\partial x_{\mathrm{A}} \partial x_{\mathrm{B}}} \\
\frac{\partial^2 \hat{G}_{\mathrm{mix}}}{\partial x_{\mathrm{B}} \partial x_{\mathrm{A}}} & \frac{\partial^2 \hat{G}_{\mathrm{mix}}}{\partial x_{\mathrm{B}}^2}
\end{bmatrix}
\right) = 0.
\end{equation}

We calculate the derivatives and determinant symbolically on a dense ternary grid ($N=1000$) for each temperature $T$ and pressure $P$, deriving the spinodal curves for each set of conditions. 
The resulting spinodal curves delineate the spontaneous phase separation boundaries in the ternary system. Compositions within the spinodals are unstable with respect to spontaneous decomposition into two phases.  The binodal, or coexistence curve, lies outside the spinodal, and the region of complete miscibility lies beyond the binodal.  

\subsection{Binodals}
The binodal pairs along the coexistence curves in ternary composition space at specified $T$ and $P$ are defined by applying the criteria for common tangents, satisfying the requirement that $\mu_i^{\alpha} = \mu_i^{\beta}$ where $\mu_i^{\alpha}$ is the chemical potential for species $i$ in phase $\alpha$ and $\mu_i^{\beta}$ is the chemical potential for that same component in phase $\beta$.  The mixing chemical potentials are related by the first derivative of the free energy of mixing

\begin{equation}
\begin{aligned}
\mu_{i,{\rm mix}}^{\alpha} - \mu_{j,{\rm mix}}^{\alpha} = \frac{\partial \hat{G}_{\mathrm{mix}}}{\partial x_{\mathrm{i}}^{\alpha}}
\end{aligned}
\label{eqn:dGdx}
\end{equation}
for independently variable component $i$ and dependent component $j$ in phase $\alpha$ and similarly for phase $\beta$. 

We found it advantageous to calculate the binodal curves themselves using a convex hull approach \citep[e.g.,][Dullemond \& Young in prep.]{Bartel_2021, Rossignol_2024}.  We calculate the lower facets of a convex hull for the free energy surface in ternary space.  The facets are triangles (2-dimensional simplices) that contact the free energy surface. Binodes lie at the facet vertices.  The vertices faithfully map the binodal with greater precision than applying Equation \ref{eqn:dGdx} directly in cases where the curvature of the free energy surface makes the latter numerically challenging. The approaches agree where curvatures are numerically robust.     

\section{Results}
\label{Results}
\subsection{Phase changes with P and T}

\begin{figure}
\centering
   \includegraphics[width=0.5\textwidth]{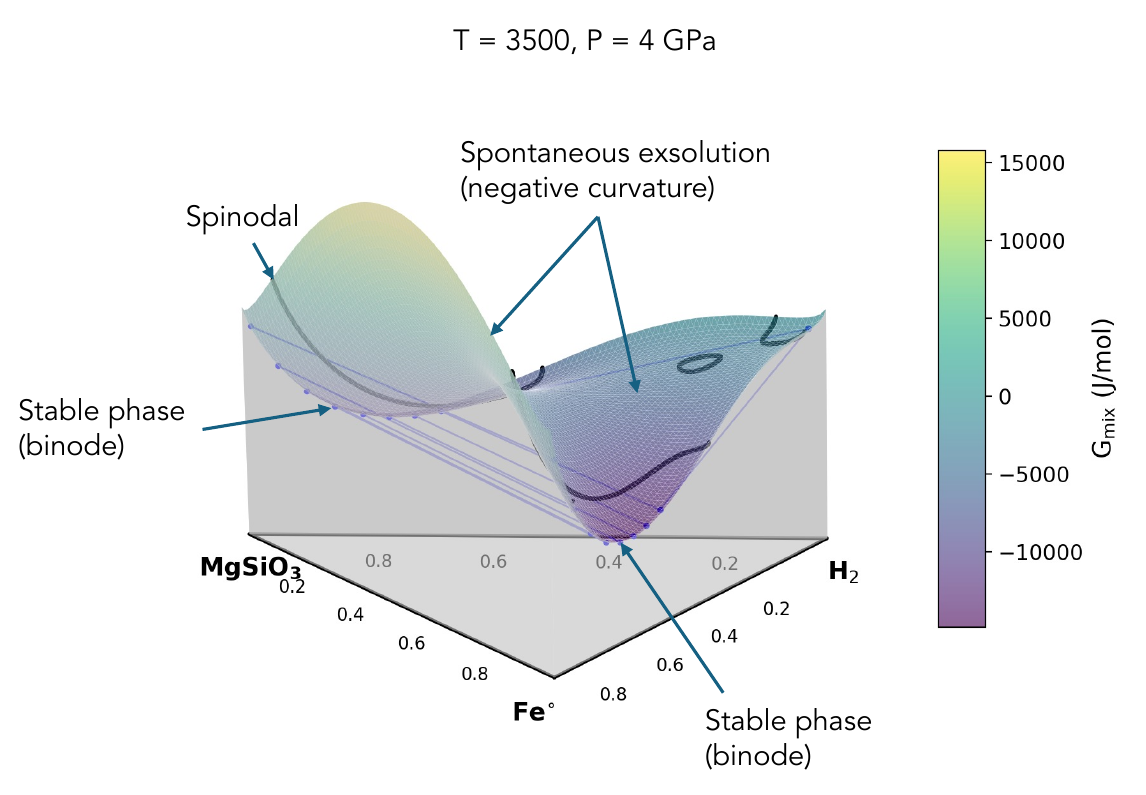}
    \caption{Gibbs free energy of mixing surface, $\hat{G}_{\rm mix}$, in the ternary system MgSiO$_3$-Fe-H$_2$ for $T = 3500$ K and $P = 4$ GPa. The heavy black lines show the spinodals denoting the regions of spontaneous decomposition into two phases. Points comprising the binodals are shown in blue connected by tie lines. The blue binodal points define the coexistence curves. Between the binodals and spinodals lies the region of metastability for two phases.}
\label{fig:3D_example}
\end{figure}

Figure \ref{fig:3D_example} shows the resulting Gibbs free energy surface, $\hat{G}_{\rm mix}$, in the ternary system MgSiO$_3$-Fe-H$_2$ for $T = 3500$ K and $P = 4$ GPa where conditions were chosen for illustration purposes only. The ternary coordinates are the mole fractions of the three components.  The heavy black lines show the spinodals. Interior to the spinodals the curvature of $ \hat{G}_{\rm mix}$ is negative, and any infinitesimal perturbation lowers the free energy.  Here, phase separation is spontaneous.  
Each of the coexisting stable phases lies at a point in the phase diagram where the curvature in $\hat{G}_{\rm mix}$ is positive and the points have the same chemical potentials (i.e., equal $\partial \hat{G}_{\mathrm{mix}}/\partial x_{\mathrm{i}}$). These points, the binodals, are shown as blue points connected by tie lines. They define the coexistence curves for two phases in the phase diagrams. Between the binodals and spinodals lies the region of metastability for two phases.

\begin{figure}
\centering
   \includegraphics[width=0.45\textwidth]{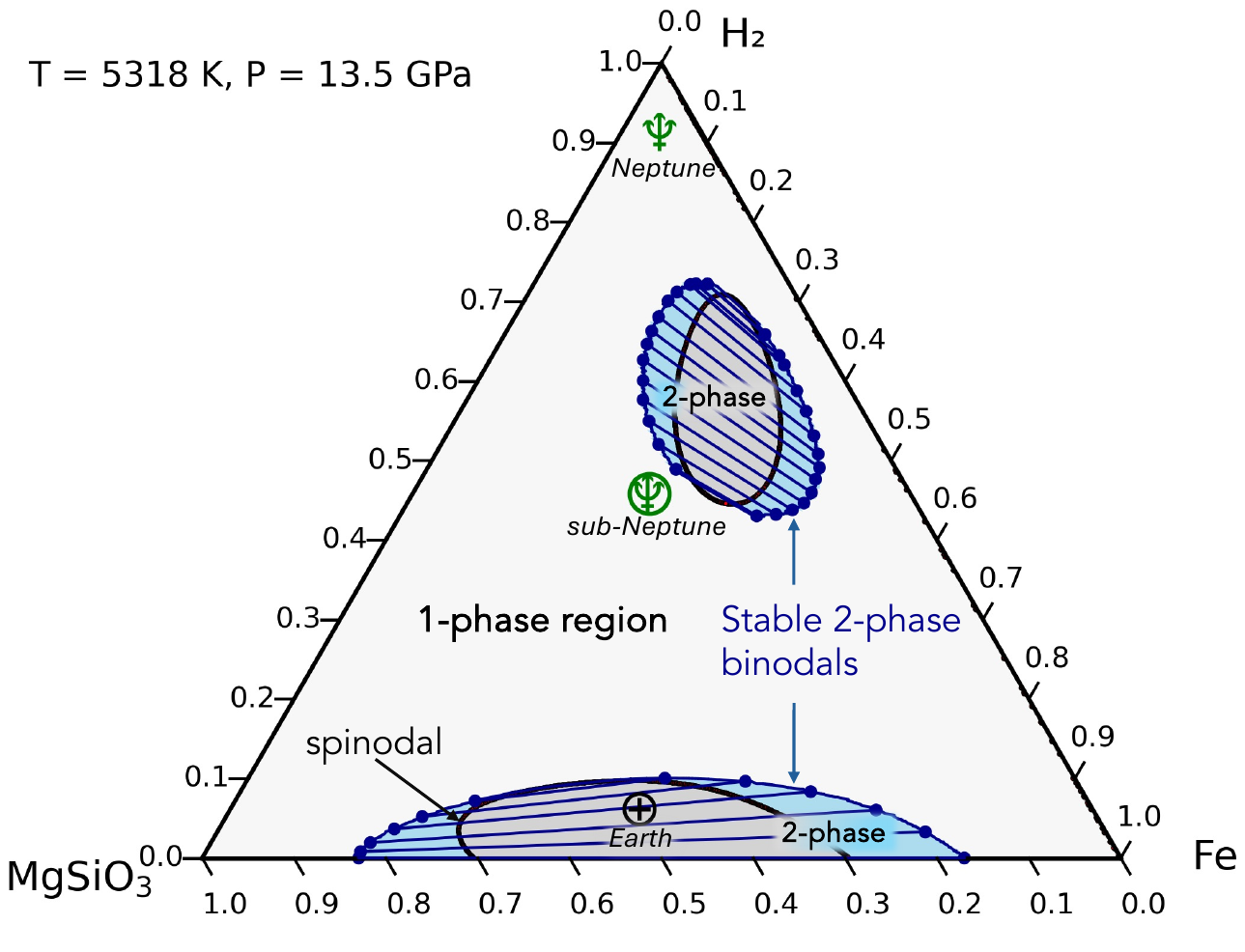}
    \caption{A ternary phase diagram for the system MgSiO$_3$-Fe-H$_2$ at $T = 5318$ K and $P = 13.5$ GPa.  Regions of 1-phase stability that lie outside of both the spinodal and binodal curves are shown in pale grey.  Regions inside of the spinodals where two liquid phases are ensured are shown in darker grey. The binodals are shown by the blue curves and coexisting phase compositions for a given bulk composition are shown by the tie lines connecting individual binodes.  Regions of two-phase metastability are between the spinodals and binodals and are shown in the sky blue color. Reference planet bulk compositions projected into this ternary space, in mole fractions, are shown for Earth ($\oplus$), our fiducial sub-Neptune (\includegraphics[height=1em]{sub-neptune.png}), and Neptune (\includegraphics[height=1em]{neptune.png}), with weight fractions of H$_2$ being 0.17\%, 2\%, and 20\%, respectively.   }
\label{fig:ternary_example}
\end{figure}

An example of an isothermal, isobaric phase diagram in ternary space is shown in Figure \ref{fig:ternary_example}. The diagram is a projection of the free energy surface in the system MgSiO$_3$-Fe-H$_2$ at $T = 5318$ K and $P = 13.5$ GPa, where the $T$ and $P$ were again chosen for illustration purposes. Ternary coordinates are mole fractions of the three components. Regions of 1-phase stability outside of both the spinodal and binodal curves are shown in pale grey. Regions inside of the spinodals are where two liquid phases must occur and are shown in darker grey. The binodals are shown by the blue curves and the compositions of coexisting two stable phases are shown by the tie lines connecting the binodes. The regions between the spinodals and binodals are where two phases might persist metastably and are shown in sky blue.  The $\oplus$ symbol represents the bulk composition of Earth assuming $0.5$ weight percent H$_2$ in its metal core \citep{Young_Nature_2023}. The \includegraphics[height=1em]{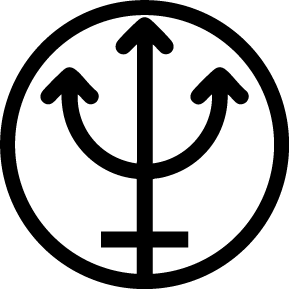} symbol represents a fiducial sub-Neptune composition for a bulk H$_2$ concentration of 2\% by mass and the remainder of the planet composed of MgSiO$_3$ silicate and Fe in 2:1 mass proportions. The projected position of Neptune in this space, \includegraphics[height=1em]{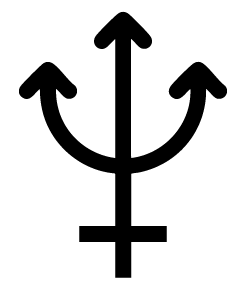}, is also shown, assuming the bulk H$_2$ concentration is 20\% by weight (approximately $0.9$ in mole fraction). Where the hydrogen-rich envelopes do not dominate the mass of H$_2$, the bulk planet and bulk interior hydrogen concentrations are similar, although not identical. 

 We constructed a new fiducial model sub-Neptune planet with a total mass of $6 M_{\oplus}$ and a bulk H$_2$ concentration of 2\% by mass based on the equations of state (EoS) for MgSiO$_3$, Fe, and H$_2$ (see Appendix B for calculation details).  In this model, the surface that separates the molten core from the outer H$_2$-rich envelope is the binodal in the system MgSiO$_3$-H$_2$. The surface of the supercritical magma ocean is therefore as described by \cite{Young_2024}. This surface moves inward as the planet cools, resulting in a lower temperature and higher pressure for the phase boundary; the conditions at the binodal separating the magma ocean from the envelope are tied to the overall thermal state of the planet. Our fiducial model was chosen to represent an evolved planet with a relatively high pressure of 4 GPa for the binodal. 

\begin{figure}
\centering
   \includegraphics[width=0.44\textwidth]{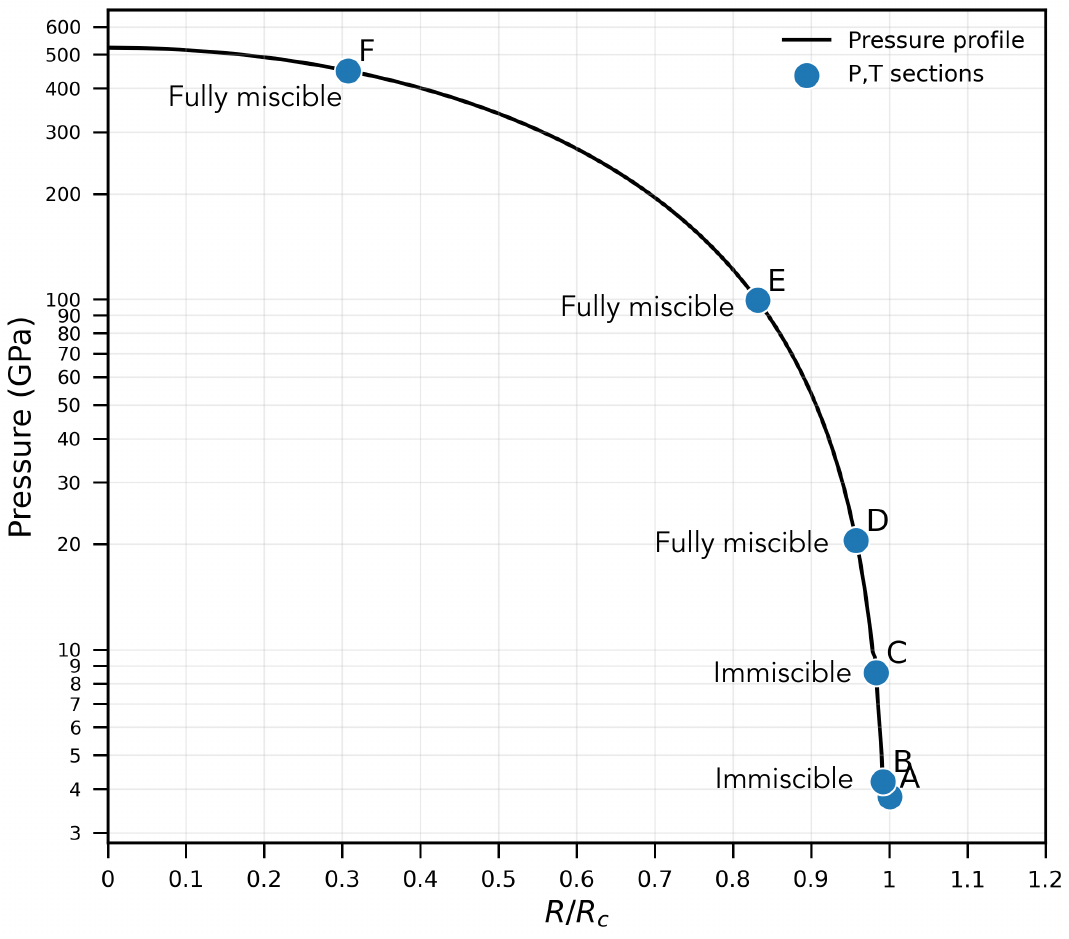}
    \caption{Pressure as a function of radial position along an isentrope in the interior of our fiducial sub-Neptune with a fully miscible core. The envelope-core interface is at $R/R_{\rm c} = 1$ where $R_{\rm c}$ is the radius of the interior exclusive of the outer hydrogen-rich envelope (the ``core"). The planet total mass is $6 M_{\oplus}$ and has a bulk H$_2$ concentration of 2\% by mass. The positions of the isobaric, isothermal phase diagrams in Figure \ref{fig:ternaries} are shown for reference. }
\label{fig:path}
\end{figure}

\begin{figure*}
\centering
   \includegraphics[width=0.95\textwidth]{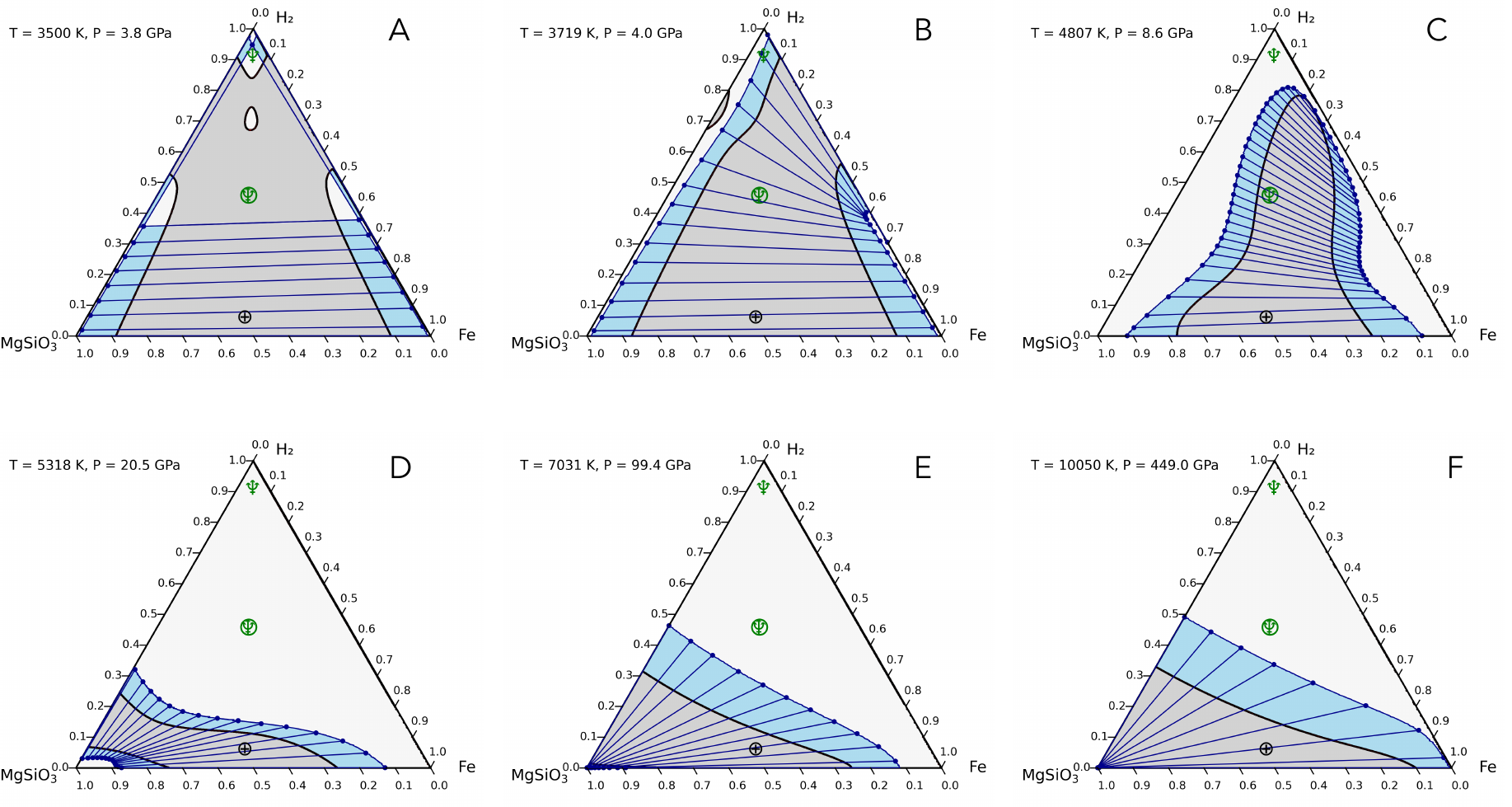}
    \caption{Isothermal, isobaric projections of the Gibbs free energy of mixing surface in the MgSiO$_3$-Fe-H$_2$ system at six different temperatures and pressures along a nominal T-P path through our fiducial sub-Neptune planet. Concentrations are in mole fractions.  The dark grey region is the region of spontaneouos exsolution into two phases and is defined by the spinodals (black curves).  The blue curves are the binodals defining the coexisting compositions for two stable phases. Example pairs of coexisting compositions are shown by the tie lines between two binodes (blue points) on the binodals.  The bulk composition of our reference sub-Neptune with 2 wt\% H$_2$ is shown by the circled Neptune symbol. The bulk composition for Earth, including H in the metal core \citep{Young_Nature_2023}, as well as the bulk composition of Neptune, are also shown for reference.  }
\label{fig:ternaries}
\end{figure*}

For these calculations, we prescribe the mass of the planet, the total H$_2$ abundance, the equilibrium temperature for the atmosphere, and the thermal state of the planet as expressed by the pressure of the binodal defining the outer boundary of the magma ocean, $P_{\rm S}$. The pressure of the binodal is a manifestation of the total energy of the planet.  Solutions are obtained by iterating between the core and atmosphere structures as follows. As a starting point for the iteration, we assume all of the H$_2$ resides in the core and calculate the surface temperature of the magma ocean, i.e. the binodal temperature where one phase becomes unstable relative to two stable phases, melt and envelope, at the prescribed concentration of H$_2$. This temperature is the reduced temperature for the adiabat in the core (i.e., the surface $T_{\rm S}$ for given $P_{\rm S}$). The shape of the binodal determines the composition of the atmosphere in equilibrium with the melt at $T_{\rm S}$ and $P_{\rm S}$ (Figure \ref{fig:isobaric_phase_diagram}).  From the pressure at the base of the atmosphere (the pressure of the binodal) and its composition, we calculate an atmosphere model, including its mass. The lower boundary conditions for the atmosphere are set by the magma-envelope binodal, and the upper boundary is set by the equilibrium temperature.  We then subtract the mass of hydrogen in this atmosphere from the core, and revise the core calculation, repeating for several iterations until the planet structure converges. The derived $P$ vs. radial position for this fiducial sub-Neptune is shown in Figure \ref{fig:path}.

 Using our ternary Gibbs free energy model, we constructed a sequence of isothermal, isobaric phase diagrams starting near the surface of the magma ocean where a supercritical magma ocean and a hydrogen-rich envelope coexist, and ending deep in the interior. The results are shown in Figure \ref{fig:ternaries}. The location of each of the phase diagrams in our model sub-Neptune is indicated in Figure \ref{fig:path}. Focusing on our fiducial sub-Neptune bulk composition, one sees a progression beginning with three phases being in equilibrium at $T = 3500$ K and $P = 3.8$ GPa (Figure \ref{fig:ternaries} A), a hydrogen-bearing silicate, hydrogen-rich iron metal, and a more hydrogen-rich envelope. As pressure increases deeper through the planet, the phase boundary separating the supercritical magma ocean from the envelope is breached, and our sub-Neptune interior is composed of hydrogen-rich liquid silicate and iron metal (Figure \ref{fig:ternaries} B). Once pressures exceed about 20 GPa ($T \sim 5000$ K), all compositions with greater than approximately 1\% by mass H$_2$, including the bulk composition corresponding to our fiducial sub-Neptune, are composed of one phase (Figures \ref{fig:ternaries} C and D). The transition at these conditions from two stable phases to one phase is at shallow depths in the planet, at $R/R_{\rm c} > 0.95$ (Figure \ref{fig:path}). Deeper still, the interiors of all planets composed of silicate, iron, and about 1\% or greater hydrogen by mass remain as a single phase (Figures \ref{fig:ternaries} E and F). The interiors of sub-Neptunes will have lower mass fractions of hydrogen than the bulk planet values because of the H$_2$ comprising the envelopes.  Our models show that for a bulk of 2\% by mass H$_2$, the interiors of sub-Neptunes similar to our fiducial case are expected to have roughly $1.3$\% by mass H$_2$ (Table \ref{Table}, see discussion below), placing them above the $\approx1 $\% threshold separating the two-phase region from the one-phase region at the highest pressures in Figure \ref{fig:ternaries}. 

This is the main result of these calculations; the interiors of planets with typical proportions of molten rock and iron, when  mixed with H$_2$ at roughly weight percent levels, form a single phase, and do not differentiate into a metal core and silicate mantle. Weight percent levels of total hydrogen are expected for sub-Neptunes, as described in the Introduction.    

\subsection{Sensitivity}

\begin{figure*}
\centering
   \includegraphics[width=0.95\textwidth]{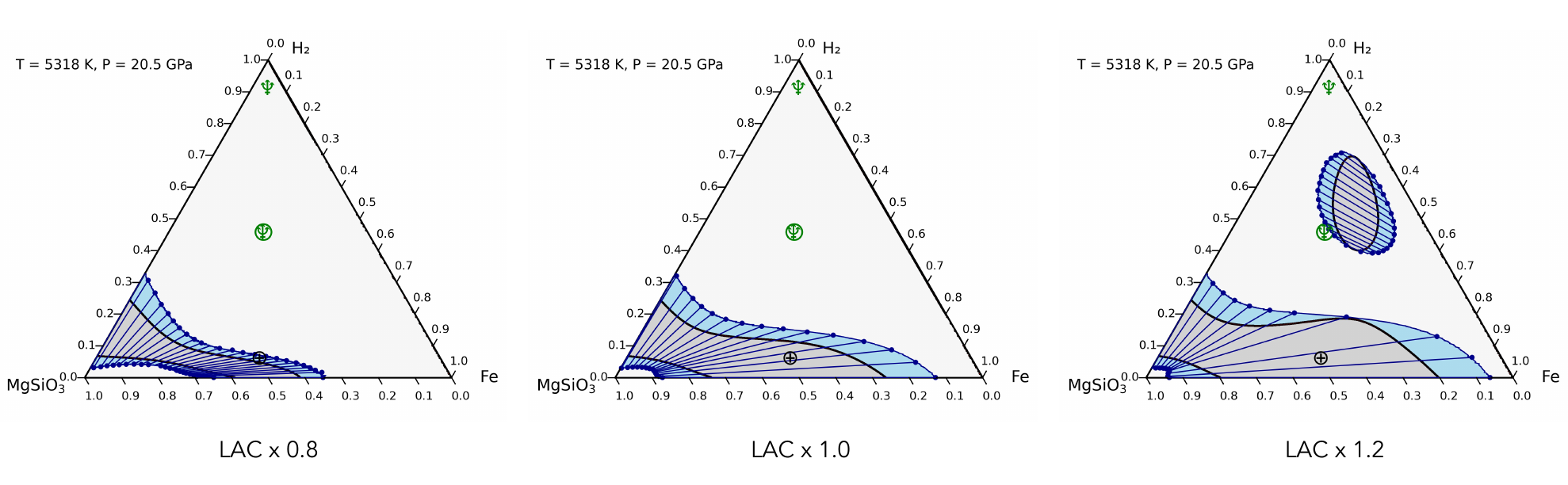}
    \caption{Changes in the topology of the ternary phase diagram shown in Figure \ref{fig:ternaries} D with variations in the interaction parameter $\rm L_{AC} = L_{CA}$ by $+20$\% and $-20$\%.  }
\label{fig:uncertainty}
\end{figure*}

It is difficult, if not impossible \citep[e.g.,][]{Gabriel_2020},  to quantify the veracity of each of the DFT-MD simulations and experiments that constrain the topology of the MgSiO$_3$-Fe-H$_2$ ternary phase diagram at the conditions relevant to sub-Neptunes. A quantitative assessment of the accuracy of the isobaric, isothermal phase diagrams is therefore not yet feasible. 

We can show, however, changes in phase diagram topologies associated with reasonable variations in the interaction parameters. Arguably, the most uncertain interaction parameter is that for the symmetrical mixing in the MgSiO$_3$-Fe binary since it has been approximated using the MgO-Fe system. \cite{Insixiengmay_2025} report a fit uncertainty of 5\% in this regular interaction parameter, but this is a measure of precision rather than potential systematic errors. We infer that the actual accuracy is likely lower by a factor of several, as is often the case in practice. Figure \ref{fig:uncertainty} shows the changes in topology in the ternary phase diagram shown in Figure \ref{fig:ternaries} D with $\pm 20$\% variations in interaction parameter $\rm L_{AC} = L_{CA}$, a spread that is four times the fit uncertainty. The changes are relatively minor.  The most significant change is that with a greater value for $\rm L_{AC}$, an island of two-phase stability appears near the composition of our fiducial sub-Neptune. Compositions that fall within this island will be composed of two phases, with their relative proportions given by the lever rule.  However, the two phases are much more similar in composition to one another than silicate mantle and iron.  One can see this by comparing the relatively short tie lines in the two-phase island to longer tie lines connecting silicate and metal in the two-phase field occupied by Earth (Figure \ref{fig:uncertainty}). The appearance and disappearance of this two-phase island is a persistent manifestation of significantly altering the Gibbs free energy surface in these calculations.  

Similarly, $\sim 20$\% variations in the other interaction parameters (and in many cases much greater variations) have relatively minor effects similar to those shown in Figure \ref{fig:uncertainty}. Among the parameters comprising the ternary mixing model, the pressure dependence of the mixing in the Fe-H$_2$ system has the most impact on the results.  For separate silicate and Fe-rich metal phases to be stable at depth in our fiducial sub-Neptune with 2\% H$_2$ (e.g., at 100 GPa along the adiabat), the pressure-dependence of the interaction parameters in Equation \ref{eqn:Fe_H2} would have to be lower than the adopted value by at least 30\%. Stabilization of the two phases results from emergence of the island of two-phase stability in the high-H$_2$ portion of the phase diagrams (e.g., Figures \ref{fig:ternary_example} and \ref{fig:uncertainty}). However, such a low pressure-dependence violates experimental and computational constraints on the relative distributions of H$_2$ between silicate and Fe metal at relevant $T$ and $P$, as described above. 

\section{Discussion}
\label{Discussion}
The complete miscibility of silicate, iron metal, and hydrogen in sub-Neptunes leads to a relatively under-dense core as compared with Earth-like layered models. As an illustration, we constructed models that can be used to compare the structures of sub-Neptunes composed of a completely miscible phase in their deep interiors with the more classical ``unequilibrated" interiors composed of an inner liquid Fe iron core and an outer liquid silicate mantle. As a predicate, we accept the premise that the interiors of sub-Neptunes should inherit significant mass fractions of hydrogen as a result of the accretion process, and that ingassing of their hydrogen-rich atmospheres after final assembly of the planets may not be the only, or even the primary, mechanism for conveying hydrogen to their interiors. Specifically, we assume that planets form from planetary embryos that accreted hydrogen-rich envelopes that in turn mixed with their molten interiors.  The likelihood that primary hydrogen-rich atmospheres intermingled with magma oceans during planet formation has been studied in the context of Earth's formation \citep{Mizuno1980, Harper_Jacobsen_1996, Jaupart_2017,Olson_Sharp_2018, Young_Nature_2023} and in the context of sub-Neptune formation and evolution \citep{Chachan2018, kite2016a, Kite2019, Schlichting_Young_2022}. The mass fraction of H$_2$ that could be accreted from a surrounding protoplanetary disk by a $0.5 M_{\oplus}$ embryo is of order 3\%  \citep{Sasaki_1990, lee2015a,  ginzburg2016a}, with more massive bodies accreting greater mass fractions of hydrogen.  The fact that sub-Neptunes do not typically have the maximum mass fractions of H$_2$ permitted by the accretion process alone speaks to the importance of various loss mechanisms during and after accretion \citep[e.g.,][]{biersteker2019a, Owen2020b}.

The model calculations are described in Appendix B, and results are summarized in Table \ref{Table}. The fully miscible model is based on the phase diagrams in Figure \ref{fig:ternaries} while the unequilibrated case ignores the phase equilibria.  For this comparison, we hold the total mass of the planet constant at 6$M_{\oplus}$ for both cases, and further specify that the total mass fraction of H$_2$ is 2\% for both, and that both have the same 1.3\% H$_2$ in their silicate-rich phases (i.e., in their mantles) in order to emphasize the effect of a metal core.  In addition, in the case of the completely miscible model, the interface between the outer envelope and the condensed supercritical core is defined by the MgSiO$_3$-H$_2$ binodal, which we use as an approximation for the ternary phase change as depicted in Figure \ref{fig:ternaries} in panels A and B. In this approximation, the H$_2$ concentrations in the magma ocean and in the outer envelope are controlled by the shape of the binary binodal (e.g., Figure \ref{fig:isobaric_phase_diagram}). For the envelope we assume that MgSiO$_3$ speciates to SiO + Mg + O$_2$ in order to achieve a realistic mean molecular weight for the atmosphere \citep{Young_2024}. We further assume that liquid in equilibrium with the gas in the envelope, as prescribed by the binodal as temperature decreases with height in the envelope, rains out immediately, rapidly distilling the envelope phase to pure H$_2$ higher up.  In the case of the ``unequilibrated" model, all of the H$_2$ in the magma ocean phase is in the silicate, with the remaining H$_2$ comprising the envelope. We specify the temperature and pressure at the magma ocean - envelope interface to be the same as that defined by the binodal in the homogeneous core case for comparison, with values of 3590 K and 4 GPa, respectively.  

\begin{deluxetable}{lccc}
\tablecaption{sub-Neptune models for planet mass of 6$M_{\oplus}$ and bulk H$_2$ concentration of 2\% by mass.}
\tabletypesize{\scriptsize}
\tablewidth{0pt}
\tablehead{
\colhead{Parameter}&
\colhead{Figure \ref{fig:planets2}A} &
\colhead{Figure \ref{fig:planets2}B} &
\colhead{Figure \ref{fig:planets1}A}
}
\startdata
$R/R_\oplus$ &3.485 &2.895 &2.618 \\
$R_{\rm core}/R_\oplus$ & 1.857 &1.988 & 1.827 \\
$M_{\rm core}/M_\oplus$ &5.960 &5.949 & 5.962\\
Metal :silicate core & 1:2 & 0 & 1:2 \\
core $\rho$ (g  $\rm cm^{-3}$)  & 5.128 & 4.175 & 5.392 \\
$T_{\rm s}$ & 3591 & 3594 & 3624  \\
$P_{\rm s}$  & 4  & 4 & 4 \\
bulk planet $\rho$ (g  $\rm cm^{-3}$) &0.781  &1.363  & 1.844  \\
mass fraction envelope & 0.655\% & 0.850\% & 0.808\% \\
$w_{\rm H_2}$ silicate core & 1.33\% & 1.34\% &1.52\% \\
$w_{\rm H_2}$ envelope & 100\% &77.78\% &76.55\% \\
\enddata
\tablenotetext{}{$T_{\rm s} = $ magma ocean surface temperature (K), $P_{\rm s} =$ magma ocean surface pressure (GPa), $w_i$ is the concentration of component $i$ by mass (\% where indicated).} 
\label{Table}
\end{deluxetable}

\begin{figure*}
\centering
   \includegraphics[width=0.95\textwidth]{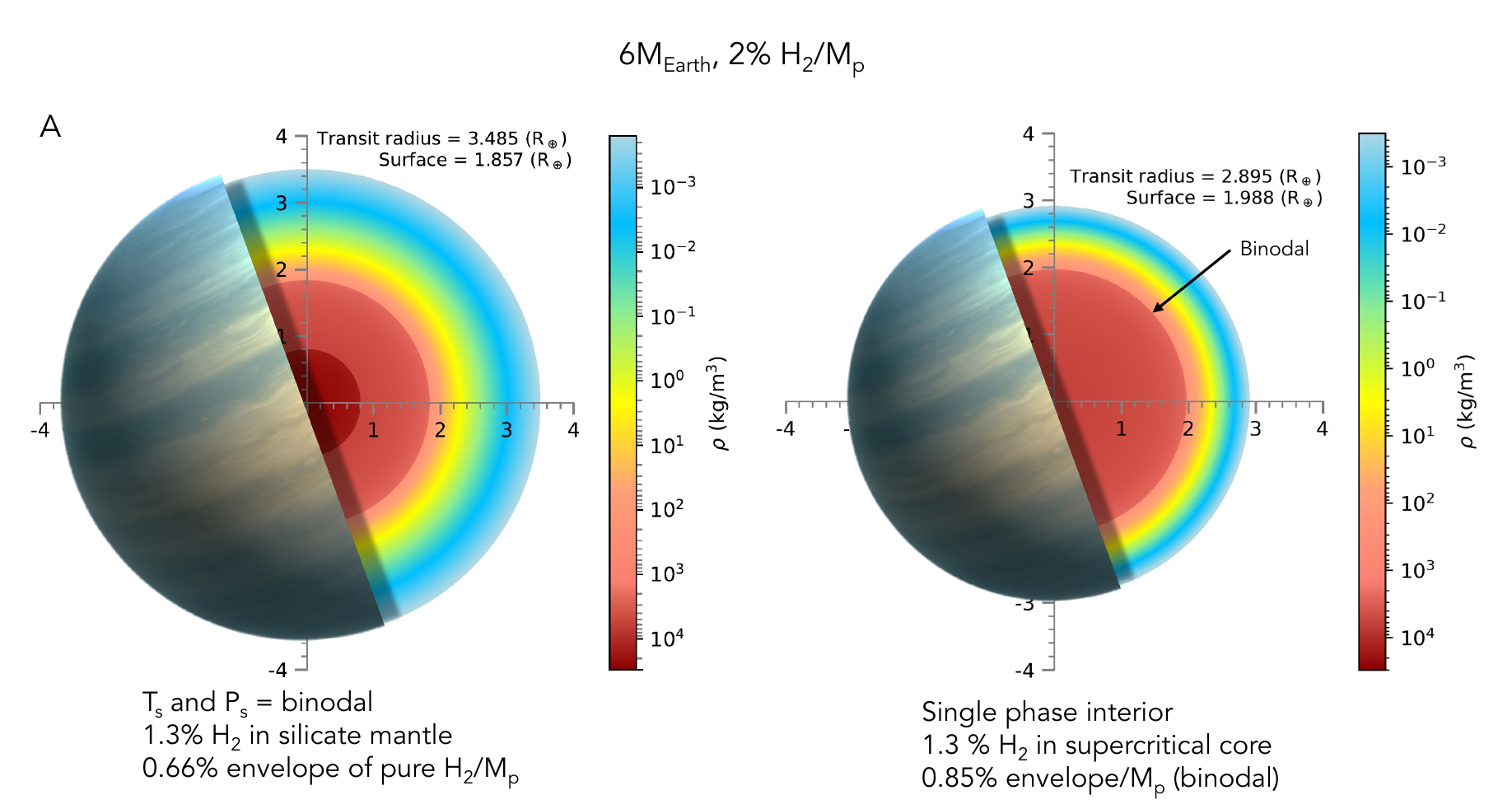}
    \caption{Comparison of two models for a fiducial sub-Neptune.  In both cases the planets have identical masses of $6M_\oplus$, equilibrium temperatures of 900K, and bulk H$_2$ concentrations of 2\% by mass. In panel A the planet is an ``unequilibrated" structure composed of a pure Fe metalic liquid core overlain by an MgSiO$_3$ melt with a specified concentration of H$_2$ to match the fully-miscible case in panel B, in turn overlain by a pure H$_2$-rich envelope.  The silicate-envelope boundary is specified to be the same $T$ and $P$ as the binodal in panel B.  The planet in panel B is composed of a supercritical magma ocean that is a fully miscible ternary mixture of MgSiO$_3$, Fe, and H$_2$. The surface of the magma ocean is defined by the MgSiO$_3$-H$_2$ binodal, where $T = 3594$ K and $P = 4.0$ GPa for the melt H$_2$ concentration of 1.3\% by mass. Note mass densities are given in kg $\rm m^{-3}$.  }
\label{fig:planets2}
\end{figure*}

Figure \ref{fig:planets2} shows the two different cases for the 6$M_{\oplus}$, 2\% H$_2$ sub-Neptune. In the ``unequilibrated" case of the core composed of pure Fe metal and a silicate mantle depicted in panel A, the bulk density of the core is 5.13 g  $\rm cm^{-3}$. The transit radius is 3.49 $R_{\oplus}$ and the radius of the core is 1.86 $R_{\oplus}$. The bulk density of the planet is 0.781 g  $\rm cm^{-3}$. In the case of the fully miscible core with an outer boundary defined by the phase change at the binodal, the core has a bulk density of 4.18 g $\rm cm^{-3}$, considerably less dense than the unequilibrated model.  In order to maintain the same imposed total mass of H$_2$ mandated by our direct comparison, the two models necessarily differ in their respective envelope mass fractions. In the unequilibrated case, the envelope comprises $0.65$\% of the planet, while in the homogeneous core case the envelope is $0.85$\% of the planet. In summary, all else equal, a planet with an entirely miscible core and an envelope that includes silicates in equilibrium with that core has a smaller radius and a less dense core than its unequilibrated counterpart planet with an envelope of pure H$_2$, a pure iron core, and H$_2$ dissolved in the silicate magma.

While useful in illustrating the large range of mass-radius relationships possible for a given total concentration of hydrogen, most studies would consider the assumption of a pure H$_2$ envelope inadequate \citep[e.g.,][]{Chen_2016,Kite2019,Bean2021,Schlichting_Young_2022,Werlen2025}.  More importantly for our purpose, the comparison above does not isolate the effect of differentiation of metal and silicate, as evident from a simple analysis of the result.  

\begin{figure*}
\centering
   \includegraphics[width=0.95\textwidth]{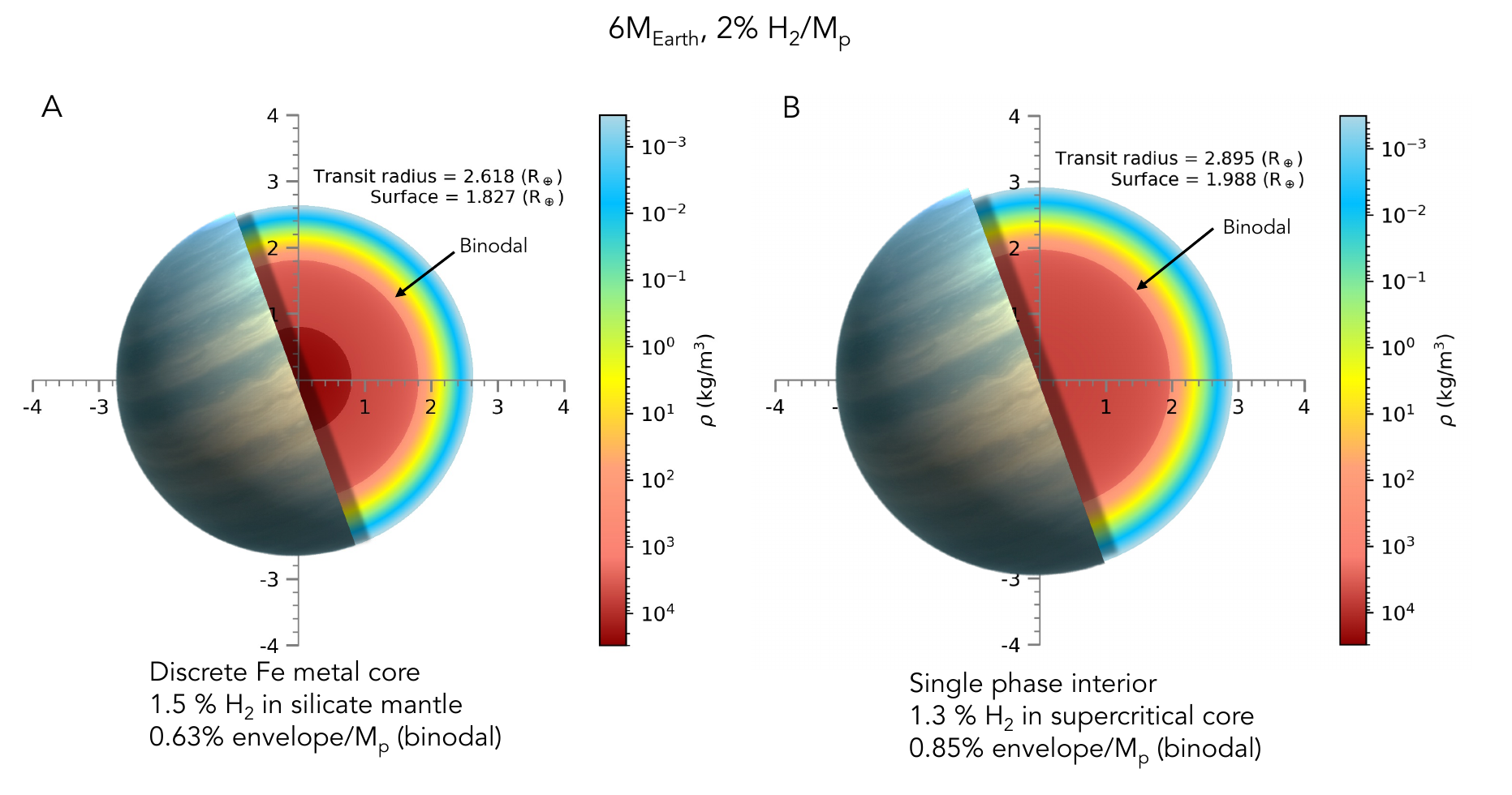}
    \caption{Comparison of two models for a fiducial sub-Neptune.  In both cases the planets have identical masses of $6M_\oplus$, equilibrium temperatures of 900K, and bulk H$_2$ concentrations of 2\% by mass. In panel A the planet is composed of a pure Fe metallic liquid core overlain by a supercritical mixture of MgSiO$_3$ and H$_2$ which is in turn overlain by an H$_2$-rich envelope.  The silicate-envelope boundary is the binodal (solvus) at a pressure of 4 GPa.  The planet in panel B is the same as in Figure \ref{fig:planets2} B and is a fully miscible ternary mixture of MgSiO$_3$, Fe, and H$_2$.    }
\label{fig:planets1}
\end{figure*}

The ratio of the core radius for the unequilibrated model in Figure \ref{fig:planets2} A to that in Figure \ref{fig:planets2} B is $0.93$ while the ratio of their transit radii is $1.19$. The mismatch is due to their different envelopes and can be understood as follows. Since the planets have the same upper atmosphere temperatures and molecular weights, the transit radii in this case depend largely on the radial thicknesses of the convective layers of the envelopes as well as on the cores.  We can understand these influences on the radii of the planets in the two models using an approximation for the radial temperature gradient due to convection for an ideal gas (taken to be the limiting case of the isentrope): 

\begin{equation}
\begin{aligned}
\nabla T=-\frac{(\gamma-1)}{\gamma} \frac{\mu_{\rm mol}}{k_{\rm B}} g ,
\end{aligned}
\label{eqn:ideal_adiabat}
\end{equation}
where $\mu_{\rm mol}$ is the mean molecular weight of the gas, $\gamma$ is the heat capacity ratio, $k_{\rm B}$ is the Boltzmann constant, and $g$ is the acceleration of gravity. The decrease in temperature from the surface to the radiative convective boundary comprising the outer radius of the convective layer of the envelope is about the same in the models in Figures \ref{fig:planets2} A and B, basically $T_{\rm s} - T_{\rm eq}$.  With the approximation of equal $dT$ values and equal heat capacity ratios, the ratio of radial thicknesses of the atmospheres, $\Delta R = R_{\rm p} - R_{\rm core}$ for the models in Figures \ref{fig:planets2} A and B, is 

\begin{equation}
\begin{aligned}
\frac{\Delta R_{A}}{\Delta R_{\rm B}} \sim \frac{\mu_{\rm mol, B}}{\mu_{\rm mol, A}} \frac{(R_{\rm core} \rho_{\rm core})_{\rm B}}{(R_{\rm core} \rho_{\rm core})_{\rm A}},
\end{aligned}
\label{eqn:approx_radii}
\end{equation}
where $\Delta R_{\rm A}$ and $\Delta R_{\rm B}$ are the thicknesses of the envelopes for the models in panels A and B in Figure \ref{fig:planets2}.  The first term on the right-hand side of Equation \ref{eqn:approx_radii} is the ratio of molecular weights.  The mean molecular weight of the envelope for the unequilibrated, pure H$_2$ envelope is, by design, $2.3$ amu (e.g., H$_2$ with an assumed He concentration) throughout, while the mean molecular weight of the envelope in the single-phase, fully-miscible interior model is $3.5$ amu near the base where much of the mass resides, transitioning  upward to 2.3 amu at lower densities.  The first term can thus be approximated as $3.5/2.3 = 1.52$.  The second term can be evaluated using the radius and density data for the cores in Table \ref{Table}, yielding a value of 0.871.  The simplified prediction for the ratio $(R_{\rm p} - R_{\rm core})_{\rm A}/(R_{\rm p} - R_{\rm core})_{\rm B}$ from Equation \ref{eqn:approx_radii} is $1.33$. The full model ratio is $1.79$ (Table \ref{Table}), the difference being attributable to the effects of convective inhibition, a moist adiabat, and other factors not included in this simple analysis. These calculations illustrate that the greater density of the core in the unequilibrated case compared with the fully miscible core should decrease the radius of the planet, all else equal, but the difference in mean molecular weights in the lower atmospheres overwhelms this effect. 

In order to better isolate the effect of the core density on the bulk density of our fiducial $6 M_\oplus$ planet, we compare the fully miscible interior model in Figure \ref{fig:planets2} B with a similar model in which the only difference is that there is a pure Fe, Earth-like metal core.  The Fe metal and silicate  are in 1:2 proportions.  Again, the total planet mass and mass fraction of hydrogen are maintained. The result is shown in Figure \ref{fig:planets1}.  Here the effects on bulk density due to  the characteristics of the differentiated core and the fully miscible core are evident.  Presence of a pure Fe core interior to the silicate increases the overall core density of the planet from $4.18$ g $\rm cm^{-3}$ to $5.39$ g $\rm cm^{-3}$. The planet with the pure Fe central core has a smaller transit radius by about 10\% compared with the fully miscible model, with values of $2.62$ $R_{\oplus}$ and $2.90$ $R_{\oplus}$ for the two bodies, respectively. The pure Fe central core case results in a greater bulk density for the planet of 1.84 g $\rm cm^{-3}$ compared with the fully miscible case where the bulk density is $1.36$ g $\rm cm^{-3}$, an increase of 35\%. 

Although the difference in transit radii is impacted by core densities, here again the envelopes also contribute. However,the important conclusion is that the envelopes enhance the effects of different core densities on the average planet density in this case, rather than opposing the effects of core densities as in the previous case (Table \ref{Table}).   

The result of this comparison of models in Figure \ref{fig:planets1} is that substituting a single, supercritical silicate-Fe-H$_2$ phase for an Earth-like central Fe metal core and an overlying H-bearing silicate magma ocean affects the observable bulk density of the planet when the total mass fraction of H$_2$ for the planet is the same, with a difference in radius of order 10\%. An even more significant change is the differences in the central core temperatures. The central temperature of the Fe-core case is about double that of the fully miscible case. The sum of thermal and gravitational potential energies for the two planets are therefore also different, by about 30\%, with values of $-3.872\times 10^{33}$ and $-4.995\times 10^{33}$ Joules for the fully-miscibile and Fe-core cases, respectively. While they are of same mass and the same total H$_2$ mass fraction, these planets would have arrived at their respective configurations along very different evolutionary paths.

\section{Conclusions}
\label{section:conclusions}
We show that based on our present understanding of the phase equilibria of the MgSiO$_3$-Fe-H$_2$ system, the interior of sub-Neptunes are unlikely to be composed of a discrete silicate mantle and iron core during, and presumably after, their magma ocean phase of evolution. Discrete silicate and metal phases are expected to occur in only the very shallowest depths of these planets.  The mixed-phase core is under-dense, with a compressed bulk density closer to 4.2 g $\rm cm^{-3}$ for a core mass of 2 $M_{\oplus}$ as compared with Earth-like compressed bulk densities of 5.4 g $\rm cm^{-3}$ that obtain for analogous planets with discrete metal cores. The boundary between this fully miscible core and the outer H$_2$-rich envelopes corresponds to the phase change from a supercritical melt to coexisting melt and envelope.  The base of the envelope is predicted to have significant fractions of heavy elements.  

These results are important for motivating investigations into the equations of state for these core materials, as well as the chemistry of sub-Neptune cores.  With our current assumptions of linear mixing of densities among the three components, and compressibilities similar to MgSiO$_3$, the effect of the supercritical, one-phase cores on the bulk density of the planets is on the order of 30\%. We have shown that miscibility is a mechanism for storing hydrogen throughout the cores of these planets that is distinct from extrapolations of solubilities of gases into silicate melt. The physical chemistry is much richer, and deserving of further exploration if we are to understand this important class of planets. With this in mind, it is tempting to extrapolate these results to the more massive Neptune-like planets as well; might ice giants contain within them supercritical mixtures dominated by Mg, Si, O, Fe and H? We are not aware of any data nor strong arguments against such a proposition.

\section{Acknowledgements} 
We acknowledge financial support from NASA grant 80NSSC24K0544 (Emerging Worlds program).  

 \vspace{20pt}

\section*{Appendix A: DFT simulations}

We use density functional theory-molecular dynamics (DFT-MD) to validate mixing along the binary joins involving silicate.  DFT is based on the concept that the Schrödinger equation may be reformulated so that the charge density, rather than the total wavefunction, is the central quantity \citep{Hohenberg1964,Kohn1965}.  We follow the methods described by \cite{Insixiengmay_2025}, \cite{Gupta2024_water}, and \cite{Stixrude2025}.  Briefly, we use DFT to determine the ground-state electron densities and energy states of the systems. The Kohn–Sham potentials and orbitals (wavefunctions) are represented with a complete, orthogonal set of basis functions using the Projector Augmented Wave (PAW) method, as implemented in the VASP code \citep{Kress1996,Kress_Joubert1999}.

Simulations are performed in the canonical ensemble in which the total number of particles, the volume, and the temperature are held fixed \citep{Hoover1985,Nose1984}. At each time step, the atomic coordinates, momenta, and forces are recorded, as are the internal energy and stress tensor computed via DFT.  

The methods for the two-phase simulations are as described by \cite{Stixrude2025}, \cite{Gupta2024_water}, and \cite{Insixiengmay_2025}. The simulations are initiated in the following manner.  We first perform one-phase simulations on each of the two pure phases.  The volumes of the periodic simulation cell are identical in the two phases and the number densities are chosen so that the equilibrated pressure is identical in the two phases.  We then initiate a two-phase simulation by constructing a supercell, consisting of half one phase and half the other.  We then allow the atoms to move in accordance with the DFT forces acting on them.  Mass exchange occurs spontaneously according to the chemical driving forces manifested in the DFT forces acting on the atoms.

For MgSiO$_3$-H$_2$ system, solid MgSiO$_3$ comprising 200 atoms and was run for 5 picoseconds at 6000 K in order to ensure that the solid structure melted. It was then cooled to the initial temperatures for the simulations and allowed to equilibrate for an additional 3-5 picoseconds. The same equilibration was performed for the pure H$_2$ phase, consisting of 176 atoms, prior to joining the two phases as shown in Figure \ref{fig:4000K_miscibility}.    

Over the course of the simulations, the two-phase system evolved to establish a dynamic equilibrium such that the composition of the two phases was no longer changing with time. The higher temperature 4,000 K simulation ran for 13.5 picoseconds and the lower temperature 3,000 K simulation ran for 16.5 picoseconds. The composition of each phase was monitored by tracking the concentration of each atom type in each phase. 

By the end of the simulations, the 4,000 K system  evolved into a single phase. The weight percent of H across the entire simulation reached 4.3\% - which is the weight percent of H in the entire system. The 3,000 K simulation remained immiscible for the duration of the simulation. The weight percent of H in the silicate-rich phase never exceeds 1\%, while it remained the dominant species in the H-rich phase.  

The methods for the MgSiO$_3$-Fe system were essentially the same, where both phases were heated to melting, then equilibrated in $P$ and $T$ prior to the initiation of mixing. The model consists of 18 Mg atoms, 18 Si atoms, 54 O atoms, and 76 Fe atoms. We checked the convergence of the results in Figure \ref{fig:9000K_miscibility} by decreasing time steps, settling on steps of $0.5$ fs. Fe atoms are non-spin polarized in these calculations. Band gaps are therefore not properly modeled, but we do not anticipate that this simplification changes the basic result of miscibility. See \cite{Insixiengmay_2025} for an assessment of the effects of spin polarization on miscibility in the MgO-Fe system. Future models should include spin polarization.

\section*{Appendix B: Planet models}

{\bf Core structure:} To derive the core structure, we solve the system of equations \citep[e.g.,][]{seager2007a}:
\begin{equation}
    \frac{dm}{dr}=4 \pi r^2 \rho,
\label{eq:dmdr}
\end{equation}
\begin{equation}
    \frac{dP}{dr}=-\frac{Gm \rho}{r^2},
\label{eq:dPdr}
\end{equation}
and,
\begin{equation}
\left(\frac{dT}{dr}\right)_S=-\gamma(\eta)\frac{T}{P}\rho g,
\label{eq:dTdr}
\end{equation}
where $m$ is the mass contained within radius $r$, $\rho$ is mass density, and $\gamma(\eta)$ is the Gr\"uneisen parameter that varies with $P$ and $T$ through the compressibility factor  $\eta = \rho/\rho_0$.   Here we have assumed the core is fully convective, with a limiting isentropic $dT/dP$ gradient. 
Numerically integrating Equations (\ref{eq:dmdr}) through (\ref{eq:dTdr}) for core and mantle yields a density and temperature profile for the core.

{\bf Material properties of the core:}  For liquid silicate densities, we use an equation of state  fit to the MgSiO$_3$ liquid properties determined by \cite{DeKoker2009} using the algorithms of \cite{Wolf_2018}. Total pressure is computed by combining the elastic (cold) compression term from a Vinet equation of state with thermal pressure contributions, following the formulation of \citet{Wolf_2018}. The thermal pressure corrections consist of vibrational energy contributions ($\Delta P_\mathrm{E}$) and deviations along the isentrope ($\Delta P_\mathrm{S}$). These corrections are calculated self-consistently as functions of the compressibility factor $\eta = \rho/\rho_0$, temperature $T$, and volume $V$. The effective heat capacity $C_{\rm P}$ also varies with the compressibility factor $\eta$ along the adiabat.

 For liquid iron, we use a Vinet equation of state modified to include thermal pressure from  \cite{Kuwayama2020}. The Gr\"uneisen parameter is a simple function of the compression ratio, which in turn determines the thermal contribution to pressure.  

{\bf Mixing:} For regions of the planet inside the binodal, hydrogen, Fe, and MgSiO$_3$ melt are completely miscible and form a homogenous mixture. We assume that the EoS for the mixture of MgSiO$_3$ and H$_2$ is the same as that for MgSiO$_3$, with adjustment for the effect of H$_2$ on the density, as described below. We include the effects of Fe by mixing compressed Fe and the MgSiO$_3$-H$_2$ mixture linearly, also  described below.  

The presence of hydrogen in the silicate melt reduces its density. Our DFT-MD simulations show that at 6000 K and 3.5 GPa, the density of the mixed phase decreases from 2.5 g $\rm cm^{-3}$ for pure MgSiO$_3$ to 1.35 g $\rm cm^{-3}$ for the mixture of MgSiO$_3$ and 4\% H$_2$. We find that a linear mixture of the compressed densities of MgSiO$_3$ and H$_2$ by volume reproduces this shift in density at these conditions.  We include the effects of H$_2$ on the density of the supercritical melt by calculating the density of the mixture at the pressures and temperatures of the surface of the magma ocean, $\rho_0$, as these conditions resemble those for the simulations. The density of the mixture is obtained using 
\begin{equation}
    \frac{1}{\hat{V}_{\rm mix}} = x_{\text{H}_2}\frac{1}{\hat{v}_{\text{H}_2}} + x_{\text{sil}} \frac{1}{ \hat{v}_{\text{sil}}},
\label{eq:Vmix}
\end{equation}
where $\hat{v}_{\text{H}_2}$ and $\hat{v}_{\text{sil}}$ are the molar volumes, and where $x_{\text{H}_2}$ and $x_{\text{sil}}$ are the mole fractions of hydrogen and silicate in the binary mixture. Densities are then MW$_{\rm mix}/\hat{V}{\rm mix}$ where MW$_{\rm mix}$ is the molecular weight of the mixture. Molar volumes are obtained from densities and the equations of state using $\rho_i/{\rm MW}_i$. This formulation is consistent with the assumption that addition of H does not greatly affect the equation of state.  We fixed the densities to be $0.09$ g $\rm cm^{-3}$ and $2.5$ g $\rm cm^{-3}$ for H$_2$ and silicate, respectively, at the surface of the magma ocean. These values come from the equations of state for hydrogen from \cite{Chabrier2019} and our {\it ab initio} molecular dynamics simulations of MgSiO$_3$ melt at 6000 K and 3.5 GPa.   

\revisedtext{We model Fe dissolved in a supercritical silicate--H$_2$ melt by assigning Fe a partial molar volume that reflects Fe occupation of Mg-like cation sites within the silicate melt framework. We let $\phi\in[0,1]$ denote the fraction of the pure MgSiO$_3$ formula-unit molar volume attributable to the Mg site, and let $\alpha=(r_{\mathrm{Fe}}/r_{\mathrm{Mg}})^3$ scale these site volumes by the ratio of octahedrally-coordinated ionic radii \citep{Shannon_1976}. If $\bar V_{\mathrm{MgSiO_3}}^{\mathrm{pure}}={\rm MW}_{\mathrm{MgSiO_3}}/\rho_{\mathrm{MgSiO_3}}^{\mathrm{pure}}$ is the molar volume of pure MgSiO$_3$ at the local $P$ and $T$, the Fe-site  partial molar volume is
\begin{equation}
\bar V_{\mathrm{Fe\, site}}^{\mathrm{eff}}
= \phi\,\alpha\;\bar V_{\mathrm{MgSiO_3}}^{\mathrm{pure}}(P,T),
\end{equation}
and the corresponding effective partial Fe density is
\begin{equation}
\rho_{\mathrm{Fe\, site}}^{\mathrm{eff}}= \frac{{\rm MW}_{\mathrm{Fe}}}{\bar V_{\mathrm{Fe\, site}}^{\mathrm{eff}}}.
\end{equation}
Assuming additivity of specific volumes (no excess mixing volume), the density of the miscible melt with Fe mass fraction $w_{\mathrm{Fe}}$ is given by the harmonic mean
\begin{equation}
\frac{1}{\rho_{\mathrm{mix}}}
= \frac{1-w_{\mathrm{Fe}}}{\rho_{\mathrm{MgSiO_3+H_2}}}
+ \frac{w_{\mathrm{Fe}}}{\rho_{\mathrm{Fe}}^{\mathrm{eff}}},
\end{equation}
where $\rho_{\mathrm{MgSiO_3+H_2}}$ is the density of the MgSiO$_3$+H$_2$ mixture derived at $P$ and $T$. An empirical value for $\phi$ of 0.5 is obtained from the  MgO-Fe system \citep{Insixiengmay_2025}). We use the MgO-Fe rather than the MgSiO$_3$-Fe simulations because in the former case NPT ensemble simulations were used to derive volumes at constant pressure, affording the required volumetric data for mixing. In the models reported here we use $r_{\mathrm{Fe}}=0.078~\text{nm}$ (VI, high spin), and $r_{\mathrm{Mg}}=0.072~\text{nm}$.  Results are insensitive to the precise values for these constants. }

{\bf Structure of the envelope:} The structure of the envelope is obtained using a model similar to that published previously by \cite{Young_2024}. We integrate numerically the equations,

\begin{equation}
    \frac{dm_{\rm Atm}}{dr}=4 \pi r^2 \rho,
\label{eq:dmdr_atm}
\end{equation}
\begin{equation}
    \frac{dP}{dr}=-\frac{Gm \rho}{r^2},
\label{eq:dPdr_atm}
\end{equation}
and,
\begin{equation}
\frac{dT}{dr} = \min\left( \left. \nabla T \right|_{\rm rad}, \left. \nabla T \right|_{\rm conv} \right),
\label{eq:dPdr_atm_minmax}
\end{equation}

\noindent where $m_{\rm Atm}$ is the mass of the atmosphere, $m$ is the total mass contained within radius $r$, and the numerical integration is from the magma ocean outward. For the envelope equation of state (densities at $P$ and $T$) we use the tabulated function from \cite{Chabrier2019}. 

The envelope is divided into three regions. The bulk of the lower region is convective. Because the vapor and silicate melt are in equilibrium, we use the moist pseudoadiabat of \cite{Graham_2021} to evaluate the convective temperature gradient $\nabla T_{\mathrm{conv}}$. Molar heat capacities required to evaluate the pseudoadiabat from \cite{Graham_2021} are obtained from the NIST thermodynamic data base \citep{Chase1998} where we assume the vapor silicate component speciates to SiO, Mg, and O$_2$. \revisedtext{If we scale the pseudoadiabat by the adiabatic temperature gradient implied by the H$_2$ equation of state, our results presented here are virtually identical if the binodal pressure is taken to be  2 GPa instead of 4 GPa. This lower pressure for the binodal would imply a younger, less evolved planet.}  

Hindrance of convection due to the mass load of heavy elements at relatively high temperatures must be included, as described previously for similar circumstances by \cite{Misener2023}.  By including the Ledoux criterion rather than just the Schwarzschild criterion for convection, the calculations permit development of a radiative layer at the base of the atmosphere that transports the intrinsic heat coming from the magma ocean upward \citep{Guillot2010,Misener2023}. The intrinsic luminosity due to the heat emanating from the core to space becomes a determining factor for the structure of the atmosphere. Since radiative diffusion relies on the local luminosity, the temperature gradient due to radiative diffusion includes the radially-dependent luminosity, $L(r)$.  At depths well below the radiative-convective boundary, stellar radiation cannot penetrate the atmosphere.  Therefore, at these depths the radiative diffusion is given by

\begin{equation}
    \nabla T_{\rm rad}=-\frac{3 \kappa \rho L(r)}{64 \pi \sigma T^3 r^2},
    \label{eq_nablarad_r}
\end{equation}
where $L(r)$ is taken to be the intrinsic luminosity, $L_{\rm int}$. The intrinsic luminosity is manifested by the radiation temperature at the top of the magma ocean,  $L_{\rm int}=\sigma (T_{\rm int})^4 4\pi R_{\rm s}^2$. The intrinsic temperature $T_{\mathrm{int}}$ is defined by the net upward component of the flux corresponding to the surface temperature of the magma ocean as prescribed by the local opacity.  The surface temperature  and optical depth at the surface therefore define $T_{\rm int}$ according to \citep[e.g.,][]{Heng_2014}

\begin{equation}
T_{\mathrm{int}} = T_{\rm s} \left( \frac{4}{3 (\tau_{\mathrm{s}} + 1)} \right)^{1/4}
\end{equation}
where using 1 instead of $3/4$ in the denominator term is an approximation for a spherical geometry, and subscript s refers to properties at the surface of the magma ocean. The optical depth at the location is obtained using the metallicity prescribed by the mixing ratios of SiO and Mg in the envelope as prescribed by the coexistence binodal curve at those conditions. 

We use the criterion for convective inhibition given by \cite{Markham2022}, expressed  in terms of the mole fraction of heavy elements relative to H$_2$, $1-x{\rm H}_2$:

\begin{equation}
(1-x{\rm H}_2)_{\rm critical}=1/\left((\Delta \hat{H}_{\rm c}/(RT)-1)(\epsilon-1)\right)
\label{eq_Markham_criterion}
\end{equation}
where $\Delta \hat{H}_{\rm c}$ is the latent heat of condensation per mole of condensate and $\epsilon$ is the ratio of the mean molecular weight of the condensable gas  to H$_2$ where we assume MgSiO$_3$ in the gas phase instantly speciates to SiO, Mg and O$_2$. The heat of condensation is calculated as $\Delta \hat{H}_{\rm c}=$ $\hat{H}_{\rm MgSiO_3} - \hat{H}_{\rm SiO} - \hat{H}_{\rm Mg} -\hat{H}_{\rm O_2}$ where the poorly known effect of H dissolved in the condensate on the latent heat is assumed to be of minor importance.  Molar enthalpies as a function of temperature are obtained from the NIST thermodynamic data base \citep{Chase1998}.  Where the mole fraction of heavy elements exceeds $(1-x{\rm H}_2)_{\rm critical}$ the temperature gradient is given by Equation \ref{eq_nablarad_r}.

The radiative transfer attending the transition from the convective layer to the upper radiative layer in the atmosphere is approximated using Eddington's two-stream 1D solution for temperature:  

\begin{equation}
\begin{aligned}
T(r)^4 = \frac{3}{4} T_{\mathrm{int}}^4 (1 + \tau(r)) + \frac{3}{4} T_{\mathrm{eq}}^4 \left(1 - \frac{1}{2} e^{-\tau(r)} \right),
\end{aligned}
\end{equation}
where both the intrinsic heat transported from below and the stellar radiation are accounted for in one dimension.  We implement the transition from convection to radiation where the gradient in T due to radiation, given by 

\begin{equation}
\frac{dT}{d\tau(r)} = \frac{1}{4 T(r)^3} \left( \frac{3}{4} T_{\mathrm{int}}^4 + \frac{3}{8} T_{\mathrm{eq}}^4 e^{-\tau(r)} \right),
\nonumber
\end{equation}
\noindent and,
\begin{equation}
\nabla T_{\mathrm{rad}} = \frac{dT}{d\tau(r)} \frac{d\tau(r)}{dr}
\end{equation}

becomes less than $\nabla T_{\rm conv}$. The two-stream solution blends radiative diffusion where optical depths are sufficient to sustain diffusion smoothly to the optically thin upper atmosphere where there is free radiation.  

We define the radius of the planet using the chord optical depth as formulated by \cite{Guillot2010} coupled with the opacities from \cite{freedman2014a} with metallicities corresponding to the composition of the gas (essentially pure H$_2$ at relevant heights for the outer atmosphere).     

\bibliographystyle{aasjournal}
\bibliography{edy_references}

\end{document}